%% file: mainsub.tex
\documentclass[sigconf]{acmart} 
\usepackage[utf8]{inputenc}
\usepackage{dirtytalk}
\usepackage{url}
\usepackage{hyperref}
\usepackage{graphbox}
\usepackage{float}
\usepackage{multicol}
\usepackage{subcaption}
\usepackage{xcolor}
\usepackage{tabularx}
\usepackage{multirow}
\usepackage{framed}
\usepackage{booktabs}
\usepackage{pifont}
\usepackage{lscape}
\usepackage{enumitem}
\usepackage{graphbox}
\usepackage{pifont}  

\newcommand*\OK{\textcolor{gray}{\ding{51}}}
\newcommand*\XMARK{\ding{55}}

\usepackage[graphicx]{realboxes}
\usepackage[symbol]{footmisc}

\title[Staying at the Roach Motel: Cross-Country Analysis of Manipulative Subscription and Cancellation Flows]{Staying at the Roach Motel: Cross-Country Analysis of Manipulative Subscription and Cancellation Flows}
\author{Ashley Sheil}
\affiliation{\institution{Maynooth University} \country{Ireland}}
\author{Gunes Acar}
\affiliation{\institution{Radboud University} \country{The Netherlands}}
\author{Hanna Schraffenberger}
\affiliation{\institution{Radboud University} \country{The Netherlands}}
\author{Raphaël Gellert}
\affiliation{\institution{Radboud University} \country{The Netherlands}}
\author{David Malone}
\affiliation{\institution{Maynooth University} \country{Ireland}}
  
\copyrightyear{2024}
\acmYear{2024}
\setcopyright{rightsretained}
\acmConference[CHI '24]{Proceedings of the CHI Conference on Human Factors in Computing Systems}{May 11--16, 2024}{Honolulu, HI, USA}
\acmBooktitle{Proceedings of the CHI Conference on Human Factors in Computing Systems (CHI '24), May 11--16, 2024, Honolulu, HI, USA}
\acmDOI{10.1145/3613904.3642881}
\acmISBN{979-8-4007-0330-0/24/05}

\begin{document}

\input{sections/abstract}


\maketitle
\input{sections/introduction}
\input{sections/background}

\input{sections/methodology}
\input{sections/results}

\input{sections/dark-patterns}
\input{sections/discussion}

\input{sections/conclusion}
\begin{acks}
    We thank Dries Cuijpers and CHI reviewers for their useful feedback. This publication has emanated from research supported in part by Science Foundation Ireland under Grant number 18/CRT/6222.
\end{acks}


\bibliographystyle{ACM-Reference-Format}
\bibliography{bibliography}

\input{sections/appendices}

\end{document}

%% file: sections/abstract.tex
\begin{abstract}

Subscribing to online services is typically a straightforward process, but cancelling them can be arduous and confusing --- causing many to resign and continue paying for services they no longer use. Making the cancellation process intentionally difficult is recognized as a dark pattern called \emph{Roach Motel}.
This paper characterizes the subscription and cancellation flows of popular news websites from four different countries, and discusses them in the context of recent regulatory changes.
We study the design features that make it difficult to cancel a subscription and find several cancellation flows that feature intentional barriers, such as forcing users to call a representative or type in a phrase.
Further, we find many subscription flows that do not adequately inform users about recurring charges.
Our results point to a growing need for effective regulation of designs that trick, coerce, or manipulate users into paying for subscriptions they do not want.
\end{abstract}

%% file: sections/introduction.tex
\section{Introduction}

\label{sec:intro}
\begin{figure}
    \centering
    \frame{\includegraphics[width=0.9\linewidth]{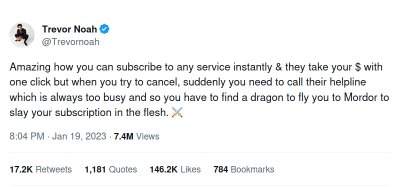}}
    \caption{A tweet posted by comedian Trevor Noah in January 2023, ridiculing the difficulties of cancelling subscriptions.}
    \label{fig:trevor-noah}
\end{figure}

In 2023, documents leaked to Business Insider suggested that Amazon intentionally designed the process of cancelling Prime subscriptions to be arduous and prolonged. The documents also revealed the company's awareness of consumers being involuntarily enrolled in and paying for their service~\cite{InsiderAmazon, FTC-Amazon-Prime-Action}.
Amazon's practices had also prompted complaints by consumer organizations in Europe~\cite{Norwegian-Prime-Cancellation-Report,EC-prime-changes-practices}. This practice of building obstacles to keep users subscribed, however, is not limited to Amazon~\cite{FTC-Vonage, FTC-ABCmouse, eHarmony-settlement, Fenty-case}. 
In fact, a 2022 survey of 2,500 Americans showed that one-third of Americans are paying for subscriptions they do not use, with 42\% feeling `locked in' to their current subscription plans \cite{bango}.
Another 2022 survey showed that consumers spend 2.5x more than they initially think on subscriptions~\cite{Tuazon2022Nov}.
Frustrated users, including celebrities (see Figure~\ref{fig:trevor-noah})~\cite{Trevor-Noah-Twitter}, turn to social media and online forums such as Reddit\footnote{The `assholedesign' subreddit popularizing manipulative designs have 3 million members, with several threads mentioning ``cancel subscription'': \url{https://www.reddit.com/r/assholedesign/search/?q=cancel\%2Fsubscription}.} to complain about these manipulative practices~\cite{nieman-lab-cancel-reasons, Consumer-Affairs-Cancel-Complaints}.
Responding to Nieman Journalism Lab's
call in 2021, several users listed unexpected credit card charges as a reason to cancel their newspaper subscriptions~\cite{nieman-lab-cancel-reasons}.
In 2022, hundreds of one-star reviews left on the customer review site ConsumerAffairs complained about subscription-related practices in industries such as dating sites, software subscriptions, radio/TV services, weight loss and health club memberships~\cite{ConsumerAffairs}. 

New laws and regulations across the globe have been recently passed or proposed to better protect consumers against unfair practices related to subscriptions~\cite{DSA-easy-cancel-press-release, FCCAreportFinland}. A notable example of these new laws is the German Fair Consumers Act, which requires a `termination button' to make it easy to cancel subscriptions \cite{Termination-Button-Germany}. In the US, the Federal Trade Commission (FTC) proposed a `Click to Cancel' option for cancelling~\cite{FTC-click-to-cancel-proposal}. Finally, the UK by 2024 aims to \say{ensure consumers can easily exit contracts without any unnecessary steps (ideally one click)}~\cite{fieldfisher-change-to-uk-sub}.

\subsection{Objectives \& Contributions}
\label{subsec:ObCon}
Motivated by user complaints, the recent and upcoming regulatory changes, and the surge of subscriptions as a business model, this paper investigates problematic practices in the context of online news subscriptions.
Existing research typically focuses on dark pattern taxonomies and/or user studies~\cite{UI-dark-patterns-and-where-to-find-them, bergman2021dark, mathur2019dark, bongard2021definitely, maier2020dark}, but the empirical research on specific dark patterns such as \emph{Roach Motel} is limited.  
In order to fill this research gap, we characterize the subscription and cancellation paths to news sites from five different locations. Specifically, we posed as a German, Dutch, English, Texan and Californian consumer and subscribed to a list of up to ten news sites for each persona's country.
In addition, our European personas also subscribed to the same American news sites to observe any differences in subscribing and cancelling based on subscribers' location.

The main contribution of this work is a systematic cross-country comparison of subscription and cancellation flow designs of popular news websites.
We study these flows through the lens of requirements imposed by the relevant existing and upcoming regulations.
In addition, we provide a comparison of subscription and cancellation flows in terms of effort necessary to achieve the two respective goals. We show that in all the jurisdictions studied, the \emph{Roach Motel}  is still present, though to varying degrees.
Finally, we contribute a dataset of over 100 screen recordings of subscriptions and cancellations\footnote{Cancellation videos of our dataset can be viewed on \url{https://github.com/roach-motel/chi-24}}. The dataset also includes emails received during or after the cancellation. 
This comprehensive dataset serves as a valuable resource for researchers studying subscription-based services from various perspectives.

%% file: sections/background.tex
\section{Background \& Related Work}
\label{sec:background}

During the COVID-19 lockdowns,  an increasing number of individuals turned to the internet for shopping, education, entertainment and connection \cite{PEW}. This coincided with a surge in the global subscription e-commerce market, which is projected to rise from \$72.91 billion in 2021 to \$904.2 billion by 2026 \cite{Forbes}.
However, not everyone is satisfied with their subscriptions. 
According to The Guardian, more than two million UK customers gave up their streaming services last year (2022) due to financial hardship \cite{UKCancelStream}.

For most subscribers, hard-to-cancel subscriptions can result in financial harm or, at the very least, frustration. Such cancellation challenges can have a more pronounced effect on individuals with mental health conditions.
A 2017 survey by Money and Mental Health Policy Institute revealed that 
individuals facing mental health challenges 
are more likely to forget to cancel their subscriptions and three times more likely to avoid cancelling when having to do so by phone \cite{MoneyMentalHealth}.
The same survey also revealed that 21\% of the participants expressed their inclination to postpone cancelling a service, despite no longer desiring it. The participants cited concerns that the company would apply pressure tactics to retain their subscription ~\cite{MoneyMentalHealth}.
Moreover, the commonly used tactic of requiring customers to contact a customer service representative to cancel their subscription can present an insurmountable challenge for users with hearing impairment~\cite{modernretail}.

Perhaps the most well-known example of easy-to-enroll but hard-to-cancel subscription practices is used by Amazon Prime, which has more than 200 million subscribers worldwide~\cite{AmazonPrime-Wikipedia}. 
The recent US Federal Trade Commission (FTC) complaint against Amazon alleges that the company knew that their customers involuntarily enrolled into Prime subscriptions and were frustrated about the arduous and perplexing Prime cancellation procedure \cite{FTC-AMAZON}.
The complaint asserts that it requires a minimum of six clicks to cancel the Amazon Prime subscription~\cite{FTC-AMAZON} and that the cancellation flow contains several dark patterns such as \emph{Forced Action}, \emph{Interface Interference}, \emph{Misdirection} and \emph{Obstruction}~\cite{gray2018dark,FTC-AMAZON}. 
In July 2022, Amazon agreed to change its cancellation process for the EU and EEA users, following a complaint made by the European Consumer Organisation (BEUC), the Norwegian Consumer Council and the Transatlantic Consumer Dialogue~\cite{EC-prime-changes-practices}. Thanks to these changes, EU and EEA users can cancel their subscriptions in two clicks with a clear and prominent cancel button--an option that was not available in the US until April 2023~\cite{FTC-AMAZON}.

Amazon's changes in their cancellation process for the EU and EEA highlight how regulations regarding subscriptions can vary between countries and jurisdictions. These differences in practices and regulations partly motivate our multi-vantage point investigation into subscription and cancellation flows.

\subsection{Related work}
\textbf{Manipulative design and dark patterns}
Dark patterns\footnote{As the term \textit{dark patterns} is officially used in the Digital Services Act (DSA)  \cite{article67-darkpatterns-DSA}, an EU regulation brought into law in November 2022 \cite{DSA-info}, we will use this term.} are defined by \citet{mathur2019dark} as \say{user interface design choices that benefit an online service by coercing, steering, or deceiving users into making decisions that, if fully informed and capable of selecting alternatives, they might not make}.
To date, there has been a growing number of papers in the area of dark pattern research~\cite{mathur2019dark,bergman2021dark,bongard2021definitely,di2020ui,maier2020dark,luguri2021shining}. \citet{mathur2021makes} review recent studies on dark patterns and identify that the literature does not have a consistent definition or singular concern on how dark patterns impact the user. 
Observing dark patterns through different disciplines, they propose a set of normative guidelines that can be used to evaluate dark patterns.

Coined by~\citet{Brignull}, the \emph{Roach Motel} dark pattern refers to a situation where signing up for a service is straightforward, but the process of discontinuing or terminating the service is challenging or complex. ~\citet{mathur2019dark} refer to this pattern as \textit{Hard to Cancel}.  
 \citet{gray2018dark} categorized 
\emph{Roach Motel} under their \emph{Obstruction} category of dark patterns, which is defined as \say{[m]aking a process more difficult than it needs to be, with the intent of dissuading certain action(s)}. \citet{chivukula2019nothing} analyzed 1002 posts from the subreddit `/r/assholedesign', where they found \emph{Obstruction} to be the second most frequent dark pattern discussed. Despite Chivukula et al.'s observation, \emph{Roach Motel} dark patterns appear to be less recognised compared to other dark patterns from a user's perspective~\cite{m2020towards}.
Finally, \emph{Preselection} is another dark pattern, where options that are favourable to the business are pre-selected, such as options to receive promotional emails or agreeing to recurring charges~\cite{mathur2021makes}.

\citet{einav2023selling} analysed credit card data used in payment for ten subscriptions, and revealed that
 subscription businesses benefit from inattentive customers. Using data from 23 million accounts, they found that users are more likely to cancel their subscriptions during the months when their cards are replaced (e.g., due to theft or loss). In a similar study, \citet{miller2023sophisticated} studied consumer \textit{inertia}, using a news site contract that automatically renews or cancels. They found that consumers anticipate their inattention and avoid subscribing to contracts which auto-renew. 
Recently, \citet{glaser2022optimal} compared the cancellation process of four major streaming platforms and found dark patterns in all platforms except Netflix. 
Glaser also performed a user study and found that users aged 50-59 had the most difficulty in unsubscribing compared to younger participants~\cite{glaser2022optimal}.

 

Another line of academic research on subscriptions focuses on consumer retention and the underlying factors that motivate individuals to subscribe to online news services \cite{fletcher2017paying,chyi2020still}. Many online articles and blogs also provide advice to news media on how to keep their churn rate low (the annual percentage rate at which customers stop
subscribing to a service) such as tracking what content their consumers are reading and identifying which consumers are at risk of not renewing their subscription~\cite{ApInstitute}.

\subsection{Updates to Regulation on Subscriptions}
\label{subsec:RegulationUpdates}

The countries involved in this study have seen some changes in regulations regarding subscriptions and dark patterns. These changes are discussed below.

\subsubsection{United States}
\label{subsec:USARegs}

In October 2021, the Federal Trade Commission (FTC) issued an enforcement policy statement on the Restore Online Shoppers' Confidence Act (ROSCA)~\cite{FTC-ROSCA}, which signalled a ramp-up of enforcement on dark patterns in subscriptions~\cite{FTC-Ramp-up}.
In 2023, the FTC announced their proposed amendment to the FTC act, which introduces three main requirements businesses must adhere to \cite{ftc-proposal-dp-sub,FTCAct}. 
The requirements are 1) clearly disclose all details about the subscription, including price, renewal dates and how to cancel; 2) obtain customer's expressed consent before charging them for products or services; 3) provide an easy and simple method to cancel subscriptions~\cite{ftc-proposal-dp-sub}.

\textbf{California} In 2017, an online cancellation provision was added to the California Business and Professions Code. In accordance with this provision, if a consumer chooses to subscribe to a service online, the business is obligated to offer an online cancellation option as well \cite{mccants2022canceling}.
On October 4, 2021, California amended the Automatic Renewal Law (ARL) of 2010~\cite{jdsupra}, which applies to all businesses that offer automatic renewal or continuous service to Californians. The new amendment, which went into effect in 2022, provides more explicit cancellation guidelines than the previous rule. 
Of the 50 US states, half have also adopted an automatic renewal law, and eight states with no existing laws are considering adopting them~\cite{ARL50states}. 

\textbf{Texas} To the best of our knowledge, Texas still lacks specific regulations on subscription services~\cite{CC-why-so-hard-to-cancel?,ARL50states}.
None of the bills introduced in 2009, 2011 and recently in 2021 (HB 2259) were enacted into law~\cite{CC-why-so-hard-to-cancel?}. The lack of regulation is indeed one of the reasons why we include Texas in this study, so we can compare the potential effects of local (as opposed to federal) regulation, or lack thereof.


\subsubsection{Europe}
\label{sec:europe}
In addition to national laws described below, EU-wide legislation, such as the Unfair Commercial Practices Directive (UCPD), also applies to subscriptions or cancellations in the e-commerce domain~\cite{UnfairCommercialPracticesEU}. The UCPD prohibits unfair commercial practices such as untruthful information or aggressive marketing techniques, and it lists ``Dark Design Patterns'' as one of the explicitly prohibited practices~\cite{UCPD-Blacklisted-Practices}. Moreover, the official guidance document on the interpretation and application of the UCPD mentions several potential infringements related to subscriptions~\cite{UCPD-Guidance}. These include ``not making it clear to consumers that they may enter into subscriptions by signing up to a free trial'' and ``[o]mitting or providing information in an unclear manner on the recurring costs of a subscription''~\cite{UCPD-Guidance}.
Note that the EU directives need to be incorporated (\textit{transposed}) into the national laws by the EU Member States. 
In the UK, the Consumer Protection from Unfair Trading Regulations 2008 (CPUT) was introduced to enact the EU UCPD. After the UK's exit from the EU, CPUT continues to be effective, with EU references removed \cite{UCPD-Brexit}.

In addition, the proposed EU Data Act requires businesses to refrain from subverting or impairing users' autonomy, decision-making or choices using dark patterns~\cite{eu-data-act}. Finally, the EU Digital Services Act, which will come into effect in February 2024, defines and prohibits dark patterns such as ``making the procedure of cancelling a service significantly more cumbersome than signing up to it''~\cite{DSA-final-text}.

\textbf{Germany}
In Germany, the new Fair Consumer Contracts Act (FCCA) came into effect in July 2022. The act states that businesses offering subscriptions to German consumers must make it possible for them to cancel subscriptions online. In addition, the act requires businesses to have a `termination' or `cancel now' button to easily cancel subscriptions~\cite{lawfed,cancelbutton,BGB-German-Civil-Code}\footnote{Prior to the FCCA, the inclusion of auto-renewal clauses in contracts allowed for the extension of the contract for an additional one-year term. Additionally, the required notice period before the end of the initial contract period could be three months~\cite{busch2022updating}.}.
This button can lead to a page where customers are requested to fill in their details, the contract they wish to cancel and reasons for cancelling~\cite{GermanyTerminationLaw,BGB-German-Civil-Code}.
This page should also incorporate a clear confirmation button, which, upon clicking, guarantees the termination of the contract. The consumer should receive a confirmation email immediately afterwards~\cite{cancelbutton}.

In 2022, the Federation of German Consumer Organisations (vzbv) examined around 840 cross-industry websites involving subscriptions to study how the FCCA was being implemented~\cite{cancelbutton}. The study revealed that 349 websites featured no cancellation button, 65 of these websites' cancellation buttons were hidden, and 38 featured invalid labels.
The vzbv released a website where consumers can report any website that does not have a compliant \textit{cancel button}~\cite{cancelbutton}.


\textbf{The Netherlands}
\label{sec:NLregs}
The Dutch Authority for Consumers and Markets (ACM) offer guidelines on the protection of the online consumer~\cite{ACM-updated-guidelines} for businesses involved in e-commerce.
With respect to subscriptions, the ACM states that businesses must provide clear information on how consumers can terminate a contract, and if a contract is initiated through a website, it should also be possible to terminate it through the same online platform~\cite{ACM-new-guidelines-cancel}. The ACM also states that
termination via the website should not involve excessive obstacles such as extensive questionnaires. 
The guidelines also suggest that marketing, design, and legal departments coordinate to comply with regulations and to adhere to the principles of \textit{fairness by design} and \textit{privacy by design}. The guidelines also highlight that deceptive design disproportionately affects certain customer groups, such as the elderly~\cite{ACM-updated-guidelines}.

\textbf{United Kingdom}
\label{sec:ukregs}
Responding to a \textit{super-complaint}\footnote{Broadly speaking, a \textit{super-complaint} is a complaint submitted by a designated UK consumer body about practices or businesses that significantly harm consumer interests~\cite{CMA-super-complaints}.}, the UK's Competition and Markets Authority (CMA) launched an investigation into \textit{subscription traps} and \textit{loyalty penalties} in 2019~\cite{loyalty-penalty-cma}. Their investigation focusing on anti-virus software and online video games resulted in Norton and McAfee committing to changes in their practices~\cite{CMA-Norton, CMA-McAfee}. These commitments included making auto-renewing contracts easier to understand and exit, and ensuring refund rights~\cite{CMA-Norton, CMA-McAfee}. Following this investigation, the CMA published guidelines (Oct 2021) on auto-renewal opt-outs and notifications \cite{CMA-antivirus-autorenew}. These guidelines, however, are not legally binding~\cite{NewCMAGuidance}.

In April 2022, reforms were announced by the UK government to address online dark patterns~\cite{gov-uk-new-rules}. The CMA will be given more power to deal with companies that break these rules. They will also be able to compensate consumers who lose out from these dark patterns without getting the courts involved~\cite{gov-uk-new-rules,gov-uk-reform-compet-policy}. In April 2023, the Digital Markets, Competition and Consumers Bill was submitted and is currently (February 2024) at the report stage in the House of Lords~\cite{UKBILL23}. 
The bill addresses auto-renewal and cancelling information, and highlights that consumers should be able to cancel online at any time. 

\subsubsection{Summary}

The above updated regulations from all jurisdictions share certain themes, with different levels of detail. In summary, consumers should be informed about the terms of a service when signing up, and businesses must not erect unreasonable barriers to cancellation. Consumers should be notified throughout the process, for auto-renewal and recurring charges. They should also be provided with a subscription receipt and cancellation confirmation. All jurisdictions highlight the importance of consumers being made aware of auto-renewal and being given the opportunity to cancel online. In short, these regulations aim to address the \emph{Roach Motel} problem and lay out the normative perspective ~\cite{mathur2021makes} that our study benefits from.

%% file: sections/methodology.tex
\section{Methodology}
\label{sec:methods}

    


As discussed in Section~\ref{sec:intro}, our main objective in this study is to characterize the subscription and cancellation pathways for news sites in the USA, UK, Netherlands, and Germany. We choose the origin locations to study practices in countries and states with different regulations. We picked news websites over other industries (e.g., streaming, games, beauty, makeup) and subscriptions types (e.g., physical) to avoid ethical problems surrounding collecting physical products and waste. News sites are widely used, reasonably priced, and available in each country, making them more feasible for our comparative study.

Several methods can be used to characterize the subscription and cancellation pathways. A potential approach involves direct inquiries to businesses regarding their practices, while another entails leveraging crowdsourcing to gather information from participants about their encounters with the studied businesses. In our study, we chose the method of employing 
personas for subscription and cancellation processes, which enabled us to directly collect the study data. This approach mitigated the potential for inaccurate or biased information stemming from businesses or participants.

\subsection{Selection of news sites}
For each country,
we first needed to compile a list of each country's news sites. 
In order to adhere to budget constraints for the project, we set a limit of ten news sites per country. The news sites were gathered from the Global Digital Subscription Snapshot (Q4 2022) report compiled by the International Federation of Periodical Publishers (FIPP) \cite{DigitalSnapshot}. Data related to all forms of media are included in this report, including a list of newspapers ranked according to the number of subscribers they have~\cite{Global-digital-snapshot}.  From this list, we took the top ten US and German newspapers. We omitted two US media outlets, which were not news related\footnote{The Weather Channel and Americas test kitchen} and replaced these with the next ranking news sites. We chose to include the newsletter subscription service Substack, because it offers subscriptions to independent journalists, subject-matter experts, and media platforms~\cite{wiki-substack}. 
There were only seven British newspapers and no Dutch newspapers in the Global Digital Subscription Snapshot report. To account for this, we picked the remainder of the newspapers from Wikipedia's lists of Dutch and British newspapers, which are ranked by circulation volume \cite{WikiDutch, WikiEnglish}. Since we focus on online subscriptions, instead of circulation volume, we ordered these newspapers by their Tranco rank and picked the top ten\footnote{Tranco website \url{https://tranco-list.eu/}}. 
Table~\ref{tab:media} provides an overview of the news sites from this compiled list.
Newspapers that are picked from Wikipedia by their Tranco rank are marked with an asterisk.

\begin{table*}
\resizebox{0.7\textwidth}{!}{%
\begin{tabular}{@{}clrrcr@{}}
\toprule
\textbf{Location} & \textbf{Organization} & \textbf{\begin{tabular}[c]{@{}r@{}}Num. of\\ Subscribers\\ (or Tranco\\ rank*)\end{tabular}} & \textbf{\begin{tabular}[c]{@{}c@{}}Subscription\\ Price\end{tabular}} & \textbf{\begin{tabular}[c]{@{}c@{}}Trial\\ Period\end{tabular}} & \textbf{\begin{tabular}[c]{@{}r@{}}Trial\\ Price\end{tabular}} \\ \midrule
\multirow{9}{*}{\textbf{\begin{tabular}[c]{@{}c@{}}United\\ States\end{tabular}}} & The New York Times & 8.3M & \$4.25/w & 1 y & \$1.00/w \\
 & Wall Street Journal & 3M & \$9.75/w & 1 y & \$2.00/w \\
 & Washington Post & 2.7M & \$12.00/m & 1 y & \$4.00/m \\
 & The Athletic & 1.1M & \$7.99/m & 1 y & \$1.99/y \\
 & Substack & 1M & \$5.00/m & - & - \\
 & Medium & 725k & \$5.00/m & - & - \\
 & The Daily Wire & 600k & \$14.00/m & - & - \\
 & Barrons & 535k & \$5.00/w & 1 y & \$1.00/w \\
 & \textcolor{gray}{LA Times}\dag & \textcolor{gray}{500k} & \textcolor{gray}{\$4.00/w} &\textcolor{gray}{ 6 m }& \textcolor{gray}{\$1.00/w} \\
 & Bloomberg Media & 375k & \$34.99/m & 3 m & \$1.99/m \\ \midrule
\multirow{8}{*}{\textbf{\begin{tabular}[c]{@{}c@{}}United\\ Kingdom\end{tabular}}} & Financial Times & 1M & £55.00/m & 1 m & £1.00/m \\
 & The Telegraph & 578k & £12.99/m & 3 m & Free \\
 & \textcolor{gray}{The Times/Sunday}\dag & \textcolor{gray}{421k} & \textcolor{gray}{£26.00/m} & \textcolor{gray}{3 m} & \textcolor{gray}{£1.00} \\ 
 & The Guardian & 420k & £3.00/m & - & - \\
 & Tortoise & 110k & £12.00/m & - & - \\
 & Mail + & 100k & £10.99/m & 3 m & £0.90/m \\
 & Spectator & 33k & £33.99/q & 1 m & Free \\
 & The Economist & (628*) & £18.90/m & 1 m & Free \\
 & iNews & (5k*) & £1.24/w & - & - \\  
 & \textcolor{gray}{The Week}\dag & \textcolor{gray}{(15k*)} & \textcolor{gray}{£38.99 /13 issues} & \textcolor{gray}{6 issues} & \textcolor{gray}{Free} \\ \midrule
\multirow{9}{*}{\textbf{Netherlands}} & Het Algemeen Dagblad & (4k*) & €2.19/w & 2 y & €1.50/w \\
 & Telegraaf & (5k*) & €3.00/w & 1 y & €1.75/w \\
 & De Volkskrant & (7k*) & €3.12/w & 2 y & €2.50/w \\
 & NRC & (9k*) & €4.38/w & - & - \\
 & Trouw & (13k*) & €3.12/w & 2 y & €2.50/w \\
 & \textcolor{gray}{Het Financieele Dagblad}\dag & \textcolor{gray}{(26k*)} & \textcolor{gray}{€52.00/m} & \textcolor{gray}{1 m} &\textcolor{gray}{€10.00/m} \\
 & Eindhovens Dagblad & (39k*) & €2.19/w & 2 y & €2.00/w \\
 & Noordhollands Dagblad & (44k*) & €3.00/w & 1 y & €1.75/w \\
 & Tubantia & (49k*) & €2.19/w & 2 y & €2.00/w \\
 & PZC & (65k*) & €2.19/w & 2 y & €2.00/w \\  \midrule
\multirow{8}{*}{\textbf{Germany}} & Bildplus & 603k & €7.99/m & 1 y & €1.99/m \\
 & WeltPlus & 200k & €9.99/m & 1 m & €1.00/m \\
 & Süddeutsche Zeitung+ & 136k & €9.99/m & 1 m & Free \\
 & Frankfurter Allgemeine & 91k & €4.95/w & 1 y & €2.95/y \\
 &\textcolor{gray}{Rheinpfalz Plus}\dag & \textcolor{gray}{47k} & \textcolor{gray}{€15.40/m} & \textcolor{gray}{1 m} & \textcolor{gray}{€1.50/m} \\
 & Heise+ & 45k & €12.95/m & 1 m & Free \\
 & Rheinische Post+ & 26k & €7.99/m & 3 m & €4.00/m \\
 & \textcolor{gray}{Allgemeine Zeitung+}\dag & \textcolor{gray}{21k} & \textcolor{gray}{€2.99/m} & \textcolor{gray}{-} & \textcolor{gray}{- }\\
 & Kieler Nachrichten+ & 20k & €9.99/m & 1 m & Free \\
 & The Pioneer & 17k & €25.00/m & - & - \\ \bottomrule
\end{tabular}%
}
\caption{The list of studied news websites, subscribers/rankings, subscription cost and trial subscription period/prices offered. \dag: News websites to which we failed to subscribe due to issues such as payments and  physical addresses. *: Tranco rank.}
\label{tab:media}
\end{table*}

\subsection{Personas and subscriptions}

For each country, we created a persona to subscribe to our compiled lists of news sites. Our Dutch, German and English persona subscribed to both their national and American news media sites. Our American personas subscribed to American news sites only. 
In addition to budget and time constraints, this decision was motivated by our assumptions that 1) fewer US residents may be interested in European news websites, and 2) European subscribers may be treated differently due to stricter regulations.
An example of this differential treatment is Amazon Prime's cancellation flow, which was redesigned as a simple two-click flow in July 2022, but only for EU consumers~\cite{gray2023temporal}. 


As mentioned in Section~\ref{subsec:USARegs}, California has implemented guidelines on cancelling subscriptions online, but Texas has no guidelines in place as yet \cite{ARL50guide}. For this reason, when subscribing to American media, we used a separate persona for California and Texas. This resulted in a total of five personas, which we created using a fake name generator website~\cite{fake-name-generator}.
Using a VPN connection, each persona signed up for subscriptions appearing to connect from their own country\footnote{We used Mullvad VPN~\cite{mullvad}.}. 
The subscriptions were paid for with a virtual credit card with a low limit, which we monitored for charges.

\subsection{Data collection protocol and evaluation parameters}
In order to consistently and efficiently collect data about subscriptions and cancellations, we developed a protocol (Appendix~\ref{app:A}) based on our pilot trials.
The protocol involves, for example, enabling screen recordings, adjusting the VPN location, and activating the separate Chrome user profile we created for each persona.
Following the protocol, we documented the subscription and cancellation processes by recording the screen and marking the evaluation parameters listed in Table~\ref{tab:eval-parameters} in a spreadsheet.
In compiling the parameters listed in Table~\ref{tab:eval-parameters}, we aimed to record as much detail as possible about each subscription and cancellation, pertaining to our objectives in Section~\ref{subsec:ObCon}. 
Moreover, we included parameters based on regulatory requirements in the studied countries such as whether a two-step termination button exists or whether users are informed about auto-renewals.
Additional parameters that we recorded but have not tabulated in the results section are listed in Appendix~\ref{app:D}.

While we tried to be as thorough as possible in compiling the evaluation parameters, we do not claim that they constitute an exhaustive list. Certain parameters such as `Num. of Clicks' for subscription and cancellation (S5 and C7) may not precisely capture the difficulties that users experience when engaging with these flows. We discuss further limitations in Section~\ref{subsec:limitations}.

\begin{table*}

\resizebox{0.8\textwidth}{!}{%
\begin{tabular}{@{}lllll@{}}
\toprule
\textbf{Code} &
  \textbf{Table} &
  \textbf{Parameter} &
  \textbf{Description} &
  \textbf{Relevance}\\ \midrule
S1 &
  3 &
  \begin{tabular}[c]{@{}l@{}}Renewal information\vspace{1.5mm}\end{tabular}  &
  \begin{tabular}[c]{@{}l@{}}Is information on renewal terms   \\ available on the website? \vspace{1.5mm}\end{tabular} 
  &\begin{tabular}[c]{@{}l@{}} Important for properly informing users.\\ Required by regulation such as: \\FTC Act (US, CA)~\cite{ftc-proposal-dp-sub},\\ UCPD (EU)~\cite{UCPD-Guidance}, \\ CMA Guidance (UK)~\cite{CMA-antivirus-autorenew}.    \vspace{1.5mm}\end{tabular} \\
S2 &
  3 &
  \begin{tabular}[c]{@{}l@{}}Email Confirmation \\(Subscription)\vspace{1.5mm}\end{tabular} &
  \begin{tabular}[c]{@{}l@{}}Did we receive an email \\ confirming subscription?\vspace{1.5mm}\end{tabular} 
  &\begin{tabular}[c]{@{}l@{}} To provide details on subscription.\\  Required by regulation such as:\\ FTC Act (US, CA)~\cite{ftc-proposal-dp-sub},\\ FCCA (DE)~\cite{cancelbutton}. \vspace{1.5mm}\end{tabular} \\
S3 &
  3 &\begin{tabular}[c]{@{}l@{}}
  Information on Cancelling\\ (Before Subscription)\vspace{1.5mm}\end{tabular}  &
  \begin{tabular}[c]{@{}l@{}}Is information available\\ on the interface \\before subscription?\vspace{2mm}\end{tabular} 
  &\begin{tabular}[c]{@{}l@{}} 
  Info before subscribing could aid in\\ decision to purchase subscription. \\  Required by regulation such as:\\ FTC Act (US, CA)~\cite{ftc-proposal-dp-sub},\\ACM Guidelines (NL)~\cite{ACM-new-guidelines-cancel}.\vspace{2mm}\end{tabular} \\
  
S4 &
  3 &\begin{tabular}[c]{@{}l@{}}
  Information on Cancelling \\(During Subscription) \vspace{1.5mm}\end{tabular} &
  \begin{tabular}[c]{@{}l@{}}Is information available \\on the interface  during\\ the subscription process?\vspace{2mm}\end{tabular} 
  &\begin{tabular}[c]{@{}l@{}}  Required by regulation such as:\\ FTC Act (US, CA)~\cite{ftc-proposal-dp-sub},\\ ACM Guidelines (NL)~\cite{ACM-new-guidelines-cancel}.  \vspace{1.5mm}\end{tabular} \\
S5 &
  3 &\begin{tabular}[c]{@{}l@{}}
  Num. of Clicks \\(Subscription)\vspace{1.5mm}\end{tabular}  &
  \begin{tabular}[c]{@{}l@{}}How many times do you need to \\ `click' to move to next page or task\\ when subscribing?\vspace{2mm}\end{tabular} 
  &\begin{tabular}[c]{@{}l@{}} Number of clicks quantify the \\ length of the subscription flow.\vspace{1.5mm}\end{tabular}\\
  S6 &
  3 &\begin{tabular}[c]{@{}l@{}}
  Required Sign-up Data \vspace{1.5mm}\end{tabular}  &
  \begin{tabular}[c]{@{}l@{}}When subscribing what data is requested? \\ Email, phone number etc.\vspace{2mm}\end{tabular} 
  &\begin{tabular}[c]{@{}l@{}} Often data requested is unecessary \\  \textit{Forced Action} (dark pattern) \\ adding to time needed to subscribe. \vspace{1.5mm}\end{tabular}\\
  \midrule
C1 &
  4 &\begin{tabular}[c]{@{}l@{}}
  Cancellation Method \vspace{2mm}\end{tabular}&
  \begin{tabular}[c]{@{}l@{}}Was the subscription cancelled  \\ online, by email, phone, chatbot or a form?\vspace{2mm}\end{tabular} 
  &\begin{tabular}[c]{@{}l@{}} 
  For usability purposes\\ what options are available?\\ Online cancellation required\\ by legislation such as:  \\ FTC Act (US, CA)~\cite{ftc-proposal-dp-sub},\\ FCCA (DE)~\cite{cancelbutton}\\ ACM Guidelines (NL)~\cite{ACM-new-guidelines-cancel}. \vspace{1.5mm}\end{tabular}\\
C2 &
  4 &\begin{tabular}[c]{@{}l@{}}
  Exit Survey \vspace{1.5mm}\end{tabular} &
  \begin{tabular}[c]{@{}l@{}}When cancelling, is there an exit \\ survey asking for cancellation reason?\vspace{1.5mm}\end{tabular} 
   &\begin{tabular}[c]{@{}l@{}} Exit surveys can add \\ extra steps to cancelling,\\ ACM Guidelines (NL)~\cite{ACM-updated-guidelines}.  \vspace{1.5mm}\end{tabular} \\
C3 &
  4 &\begin{tabular}[c]{@{}l@{}}
  Mandatory Exit Survey \vspace{1.5mm}\end{tabular}&
  Is the exit survey mandatory to answer?\vspace{2mm} 
   &\begin{tabular}[c]{@{}l@{}} Forcing people to take a survey\\ can constitute a barrier,\\ ACM (NL) guidelines~\cite{ACM-updated-guidelines} . \\ Dark pattern: \textit{Forced Action}.  \vspace{1.5mm}\end{tabular} \\
C4 &
  4 &\begin{tabular}[c]{@{}l@{}}
  Special Offers \vspace{1.5mm}\end{tabular}&
  \begin{tabular}[c]{@{}l@{}}Are there special offers or\\ discounts while cancelling?\vspace{2mm}\end{tabular} 
   &\begin{tabular}[c]{@{}l@{}} Offers and discounts can add \\ extra steps to cancelling.  \vspace{1.5mm}\end{tabular} \\
C5 &
  4 &\begin{tabular}[c]{@{}l@{}}
  Email Confirmation \\(Cancellation)\vspace{1.5mm}\end{tabular} &
  \begin{tabular}[c]{@{}l@{}}Did we receive an email confirming \\ cancellation?\vspace{2mm}\end{tabular} 
   &\begin{tabular}[c]{@{}l@{}} Provides proof of cancellation. \\ Required by regulation such as:\\ FTC Act (US, CA)~\cite{ftc-proposal-dp-sub},\\ FCCA (DE)~\cite{cancelbutton}. \vspace{1.5mm}\end{tabular}\\
C6 &
  4 &\begin{tabular}[c]{@{}l@{}}
  Two-step Termination \vspace{1.5mm}\end{tabular}&
  \begin{tabular}[c]{@{}l@{}}Is there a termination button such as the \\ German FCCA law stipulates, which allows \\ cancelling in two steps?\vspace{1.5mm}\end{tabular} 
   &\begin{tabular}[c]{@{}l@{}} For comparison to\\ strictest regulation \\ (FCC Act~\cite{lawfed,cancelbutton,BGB-German-Civil-Code}). \vspace{1.5mm}\end{tabular} \\
C7 &
  4 &\begin{tabular}[c]{@{}l@{}}
  Num. of Clicks \\(Cancellation)\vspace{1.5mm}\end{tabular} &
  \begin{tabular}[c]{@{}l@{}}How many times do you need\\ to `click' to move to next page or \\task when cancelling?\vspace{1.5mm}\end{tabular} 
   &\begin{tabular}[c]{@{}l@{}} Number of clicks quantify the \\ length of the cancellation flow. \vspace{1.5mm}\end{tabular} \\
\bottomrule
\end{tabular}%
}
\caption{Descriptions of evaluation parameters considered in this study. Subscription parameters are denoted by codes starting with S, and cancellation parameters by C. The ``Table'' column shows
in which table the corresponding results can be found. Additional parameters that are recorded but not used in the results are given in Table~\ref{tab:additional-params} in Appendix~\ref{app:D}}.

\label{tab:eval-parameters}
\end{table*}


In the cases where it was necessary to cancel by phone, we took notes on the conversation; such as questions and details requested, and record the call duration. 
For the subscription and cancelling steps, we record how many `click next' buttons we encounter in the subscription and cancellation flow. 
The number of `clicks' made during subscription or cancellation flows roughly correspond to the number of steps in the process. Clicks during account sign-up, those used to select survey answers from drop-down menus, and clicks made for personalizing service preferences were all exempt from the click count parameters.

The subscription and cancellation processes started in 5 January 2023 and ended in 7 February 2023.
We spent roughly two person-months to record and code the subscriptions and cancellations.
As the study budget was limited, we kept paid subscriptions for one month only and cancelled all subscriptions before the end of the payment term, or before trials ran out, to avoid being charged for a second term. 

%% file: sections/results.tex
\section{Results}
\label{sec:results}
\subsection{Overview}

We successfully signed up for 67 subscriptions on 34 distinct news websites, for a total cost of €412.9 (\$444.9). 13 of the 80 subscriptions could not be completed due to issues such as requirements to provide real addresses, and payment issues. The LA Times was the only website for which all five personas failed to subscribe, potentially due to our use of a VPN connection.
We explain the reasons behind the remaining eight unsuccessful subscriptions for each country's websites in their individual sections below. 

\begin{figure*}
    \centering
    \frame{\includegraphics[width = 0.9 \textwidth]{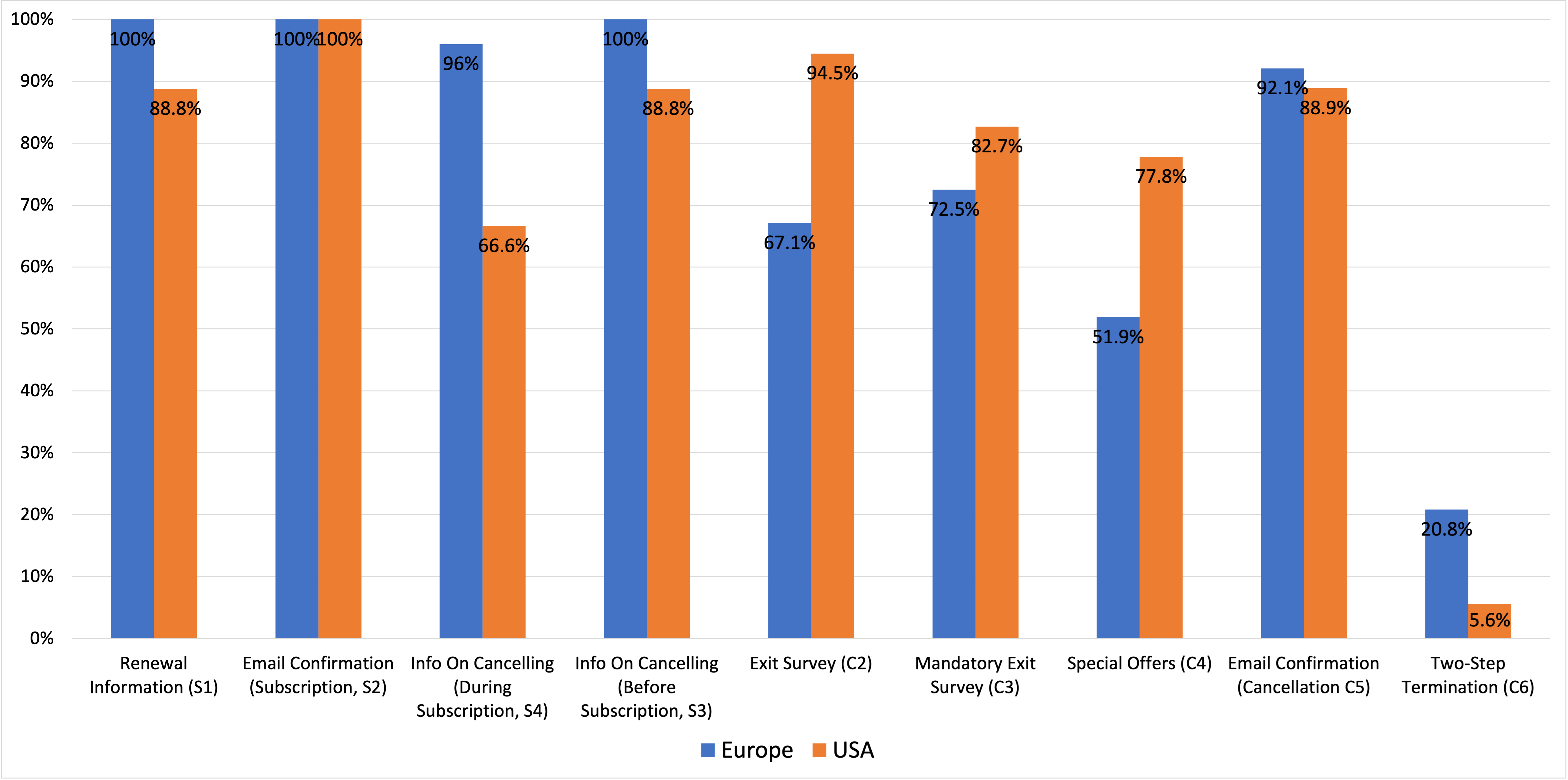}}
    \caption{Selected subscription and cancellation parameters (averaged) for European vs. American news sites. European average is taken from the averages of British persona subscribing to British sites, German persona to German sites and Dutch persona to Dutch news sites. The same applies for the US refering to American personas subscribing to American news sites.
    }
    \label{fig:EuropeVusa}
\end{figure*}

\begin{figure*}
    \centering
    \frame{\includegraphics[width=0.9 \textwidth]{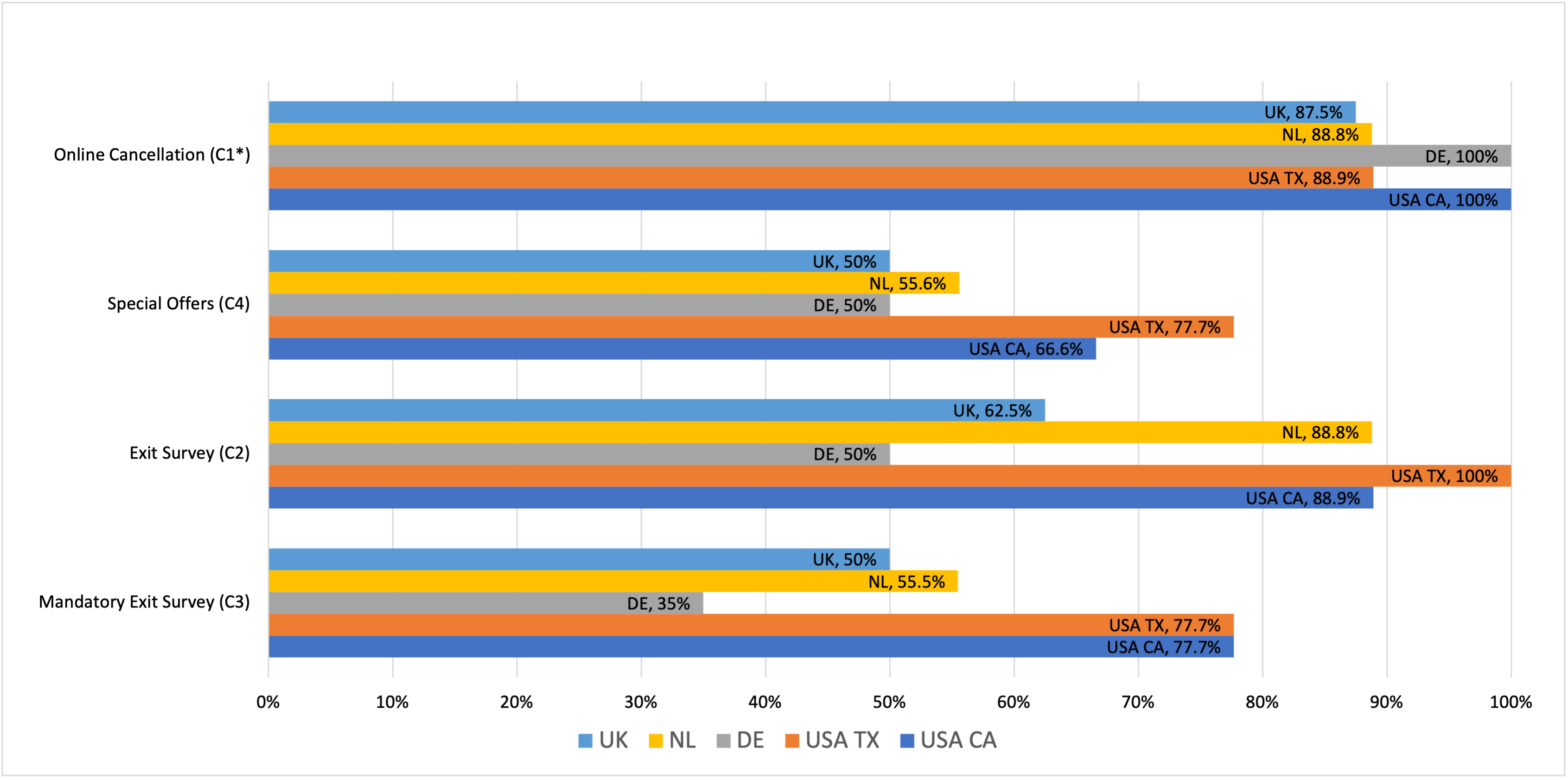}}
    \caption{Cancellation parameters averaged for each persona subscribing to sites in their own country.: Method of cancellation (C1) *Online only; if we were presented with offers to stay during cancellation (C4); if we were asked to fill an exit survey (C2); and whether the exit survey was mandatory (C3). }
    \label{fig:roachmotelstats}
\end{figure*}

Our personas were able to cancel 63 out of the 67 subscriptions online. While we acknowledge the prevalence of online cancellation as a positive outcome, we detail below various difficulties encountered during online cancellations. 
Figure~\ref{fig:EuropeVusa} displays overall differences in subscribing and cancelling between USA and Europe, with respect to parameters described in Table~\ref{tab:eval-parameters}.
According to Figure~\ref{fig:EuropeVusa}, more European websites provided information on cancellation options and automatic renewals.
They were also more likely to offer a two-step cancellation option, which is required by the German FCCA law.
On the other hand, more US websites displayed special offers and (mandatory) exit surveys during cancellation.
Other aspects of cancellation flows for each country are summarised in Figure~\ref{fig:roachmotelstats}.
Table~\ref{tab:sub-data-summary} and Table~\ref{tab:cancel-data-stats} give an overview of subscription and cancellation results.

\begin{table*}
\resizebox{0.95\textwidth}{!}{%
\begin{tabular}{@{}clccccrc@{}}
\toprule
\textbf{Country} & \multicolumn{1}{c}{\textbf{Organization}} & \textbf{\begin{tabular}[c]{@{}c@{}}Renewal \\ Info. (S1) \end{tabular}} & \textbf{\begin{tabular}[c]{@{}c@{}}Email \\ Confirmation (S2)\\ \end{tabular}} & \textbf{\begin{tabular}[c]{@{}c@{}}Info on \\ Cancelling\\  During \\ Subscription (S3) \end{tabular}} & \textbf{\begin{tabular}[c]{@{}c@{}}Info on \\Cancelling \\ Before \\ Subscription (S4)\end{tabular}} & \multicolumn{1}{c}{\textbf{\begin{tabular}[c]{@{}c@{}}Num. of \\Clicks to \\ Subscribe (S5)\end{tabular}}} & \multicolumn{1}{c}{\textbf{\begin{tabular}[c]{@{}c@{}}
Required \\Sign-up \\ Data (S6)\end{tabular}}}\\ \midrule
\multirow{8}{*}{\textbf{Germany}} & Bildplus & \OK & \OK & \OK & \OK & 3 & ENO\\
 & WeltPlus & \OK & \OK & \OK & \OK & 3 & ENO \\
 & Süddeutsche Zeitung+ & \OK & \OK & \OK & \OK & 5 & ENO \\
 & Frankfurter Allgemeine & \OK & \OK & \OK & \OK & 5 & ENO \\
 & Heise+ & \OK & \OK & \XMARK & \OK & 4& ENO \\
 & Rheinische Post+ & \OK & \OK & \OK & \OK & 4 & EO \\
 & Kieler Nachrichten+ & \OK & \OK & \OK & \OK & 5 & ENOZ \\
 & The Pioneer & \OK & \OK & \OK & \OK & 4 & ENO \\ \midrule
\textbf{Average} &  & 100\% & 100\%& 87.5\% & 100\%& 4.1 & \\ \midrule
\multirow{9}{*}{\textbf{Netherlands}} & Het Algemeen Dagblad (AD) & \OK & \OK & \OK & \OK & 3 & EN \\
 & Telegraaf & \OK & \OK & \OK & \OK & 3  & DENP\\
 & De Volkskrant & \OK & \OK & \OK & \OK & 3  & EN\\
 & NRC & \OK & \OK & \OK & \OK & 3  & AENZ\\
 & Trouw & \OK & \OK & \OK & \OK & 3  & EN\\
 & Eindhovens Dagblad (ED) & \OK & \OK & \OK & \OK & 3  & EN\\
 & Noordhollands Dagblad & \OK & \OK & \OK & \OK & 3  & EN\\
 & Tubantia & \OK & \OK & \OK & \OK & 3  & EN\\
 & PZC & \OK & \OK & \OK & \OK & 3  & EN\\ \midrule
\textbf{Average} &  & 100\% & 100\% & 100\% &100\%  & 3.0 & \\ \midrule
\multirow{8}{*}{\textbf{United Kingdom}} & Financial Times & \OK & \OK & \OK & \OK & 3  & ENOP\\
 & The Telegraph & \OK & \OK & \OK & \OK & 3  & EN\\
 & The Guardian & \OK & \OK & \OK & \OK & 2  & EN\\
 & Tortoise & \OK & \OK & \OK & \OK & 4  & EN\\
 & Mail + & \OK & \OK & \OK & \OK & 4  & DENO\\
 & Spectator & \OK & \OK & \OK & \OK & 4  & ACENOZ\\
 & The Economist & \OK & \OK & \OK & \OK & 3  & ACENOZ\\
 & iNews & \OK & \OK & \OK & \OK & 3  & EN\\ \midrule
\textbf{Average} &  &100\%  &100\%  &100\%  & 100\% & 3.3 & \\ \midrule
\multirow{9}{*}{\textbf{United States}} & The New York Times & \OK & \OK & \OK & \OK & 4  & ENOZ\\
 & Wall Street Journal & \OK & \OK & \OK & \OK & 5  & ACENOZ\\
 & Washington Post & \OK & \OK & \OK & \OK & 4  & EOZ \\
 & The Athletic & \OK & \OK & \OK & \OK & 2  & EN\\
 & Substack & \XMARK & \OK & \XMARK & \OK & 3  & E \\
 & Medium & \OK & \OK & \XMARK & \OK &3  & E \\
 & The Daily Wire & \OK & \OK & \XMARK & \XMARK &3  & ACENOZ \\
 & Barrons & \OK & \OK & \OK & \OK & 5  & ACENOZ\\
 & Bloomberg Media & \OK & \OK & \OK & \OK & 3  & CENO\\ \midrule
\textbf{Average} &  & 88.8\% & 100\% &66.6\%  & 88.8\% & 3.5 & \\ \bottomrule
\end{tabular}%
}
\caption{Overview of subscription flow features for personas subscribing to sites in their own country. United States represent both Texas and California, which had identical results for the scope of this table. Letters in the rightmost column (S6) indicate the information required during subscription---E: Email, N: Name, O: Country, Z: Zip code, D: Date of birth, A: Address, P: Phone number, C: City.}
\label{tab:sub-data-summary}
\end{table*}

\begin{table*}
\resizebox{0.95\textwidth}{!}{%
\begin{tabular}{@{}clccccccr@{}}
\toprule
\textbf{Country} & \multicolumn{1}{l}{\textbf{Organization}} & \textbf{\begin{tabular}[c]{@{}c@{}}Cancellation\\ Method (C1)\end{tabular}} & \textbf{\begin{tabular}[c]{@{}c@{}}Exit\\ Survey (C2) \end{tabular}} & \textbf{\begin{tabular}[c]{@{}c@{}}Exit\\ Survey\\ Mandatory? (C3)\end{tabular}} & \textbf{\begin{tabular}[c]{@{}c@{}}Special \\ Offers (C4) \end{tabular}} & \textbf{\begin{tabular}[c]{@{}c@{}}Post-\\ cancel\\ Email (C5)\end{tabular}} & \textbf{\begin{tabular}[c]{@{}c@{}}Two-step \\Termination (C6)\end{tabular}} & \multicolumn{1}{c}{\textbf{\begin{tabular}[c]{@{}c@{}}Num. of\\  Clicks \\ to Cancel (C7)\end{tabular}}} \\ 
\midrule
\multirow{8}{*}{\textbf{Germany}} & Bildplus & Online & \ding{108} & \ding{108} &  & \ding{108} &  & 5 \\
 & WeltPlus & Online & \ding{108} & \ding{108} & \ding{108} & \ding{108} &  & 6 \\
 & Süddeutsche Zeitung+ & Online &  &  &  & \ding{108} & \ding{108} & 5 \\
 & Frankfurter Allgemeine & Online &  &  & \ding{108} & \ding{108} &  & 7 \\
 & Heise+ & Online & \ding{108} &  & \ding{108} & \ding{108} &  & 5 \\
 & Rheinische Post+ & Online & \ding{108} & \ding{108} & \ding{108} & \ding{108} &  & 5 \\
 & Kieler Nachrichten+ & Online &  &  &  & \ding{108} & \ding{108} & 4 \\
 & The Pioneer & Online (Form) &  &  &  & \ding{108} & \ding{108} & 4 \\ \midrule
\textbf{Average} & \multicolumn{1}{c}{} &  & 50\%& *75\%& 50\% & 100\% & 37.5\% & 5.1 \\ \midrule
\multirow{9}{*}{\textbf{Netherlands}} & Het Algemeen Dagblad & Online & \ding{108} &  & \ding{108} & \ding{108} &  & 4 \\
 & Telegraaf & Online & \ding{108} &  &  & \ding{108} &  & 6 \\
 & De Volkskrant & Online & \ding{108} & \ding{108} & \ding{108} & \ding{108} &  & 5 \\
 & NRC & \textbf{Phone} &  &  &  & \ding{108} &  &  \\
 & Trouw & Online & \ding{108} & \ding{108} & \ding{108} & \ding{108} &  & 4 \\
 & Eindhovens Dagblad & Online & \ding{108} & \ding{108} & \ding{108} & \ding{108} &  & 4 \\
 & Noordhollands Dagblad & Online & \ding{108} &  &  &  &  & 6 \\
 & Tubantia & Online & \ding{108} & \ding{108} & \ding{108} & \ding{108} &  & 4 \\
 & PZC & Online & \ding{108} & \ding{108} & \ding{108} & \ding{108} &  & 6 \\ \midrule
\textbf{Average} &  & \multicolumn{1}{l}{} & 88.9\%& *62.5\%& 55.6\%& 88.9\% & 0\% & 4.9 \\ \midrule
\multirow{8}{*}{\textbf{United Kingdom}} & Financial Times & Online & \ding{108} &  & \ding{108} & \ding{108} &  & 6 \\
 & The Telegraph & Online & \ding{108} & \ding{108} & \ding{108} & \ding{108} &  & 8 \\
 & The Guardian & Online & \ding{108} & \ding{108} & \ding{108} &  &  & 5 \\
 & Tortoise & Online &  &  &  & \ding{108} & \ding{108} & 4 \\
 & Mail+ & \textbf{Phone} & \ding{108} & \ding{108} &  & \ding{108} &  &  \\
 & Spectator & Online (Form) &  &  &  & \ding{108} &  & 4 \\
 & The Economist & Online & \ding{108} & \ding{108} & \ding{108} &  \ding{108} &  & 7 \\
 & iNews & Online &  &  &  & \ding{108} & \ding{108} & 4 \\ \midrule
\textbf{Average} &  & \multicolumn{1}{l}{} & 62.5\%& *80\%& 50\%  & 87.5\% & 25\% & 5.4\\ \midrule
\multirow{9}{*}{\textbf{USA Texas}} & The New York Times & Online & \ding{108} & \multicolumn{1}{l}{} & \ding{108} & \ding{108} &  & 7 \\
 & Wall Street Journal & \textbf{Phone} & \ding{108} & \ding{108} & \multicolumn{1}{l}{} & \ding{108} &  &  \\
 & Washington Post & Online & \ding{108} & \ding{108} & \ding{108} & \ding{108} &  & 6 \\
 & The Athletic & Online & \ding{108} & \ding{108} & \ding{108} & \ding{108} &  & 6 \\
 & Substack & Online & \ding{108} & \multicolumn{1}{l}{} & \multicolumn{1}{l}{} & \ding{108} &  & 5 \\
 & Medium & Online & \ding{108} & \ding{108} & \ding{108} & \ding{108} &  & 5 \\
 & The Daily Wire & Online & \ding{108} & \ding{108} & \ding{108} & \ding{108} &  & 4 \\
 & Barrons & Online & \ding{108} & \ding{108} & \ding{108} & \ding{108} &  & 5 \\
 & Bloomberg Media & Online (\textbf{ChatBot}) & \ding{108} & \ding{108} & \ding{108} &  &  & 10 \\ \midrule
\textbf{Average} &  & \multicolumn{1}{l}{} & 100\% & *77.8\%&77.8\%  &88.9\% & 0\%& 6.0 \\ \midrule
\multirow{9}{*}{\textbf{USA California}} & The New York Times & Online & \ding{108} & \multicolumn{1}{l}{} & \ding{108} & \ding{108} &  & 7 \\
 & Wall Street Journal & Online & \ding{108} & \ding{108} & \multicolumn{1}{l}{} & \ding{108} &  & 5 \\
 & Washington Post & Online & \ding{108} & \ding{108} & \ding{108} & \ding{108} &  & 6 \\
 & The Athletic & Online & \ding{108} & \ding{108} & \ding{108} & \ding{108} &  & 6 \\
 & Substack & Online &  & \multicolumn{1}{l}{} & \multicolumn{1}{l}{} & \ding{108} &\ding{108}  & 3 \\
 & Medium & Online & \ding{108} & \ding{108} & \ding{108} & \ding{108} &  & 5 \\
 & The Daily Wire & Online & \ding{108} & \ding{108} & \ding{108} & \ding{108} &  & 4 \\
 & Barrons & Online & \ding{108} & \ding{108} & \ding{108} & \ding{108} &  & 5 \\
 & Bloomberg Media & Online (\textbf{ChatBot}) & \ding{108} & \ding{108} & \ding{108} &  &  & 10 \\ \midrule
\textbf{Average} &  & \multicolumn{1}{l}{} & 88.9\% & *87.5\%&77.8\%  &88.9\% & 11.1\%& 5.7 \\ 
\bottomrule
\end{tabular}%
}
\caption{Overview of the cancellation flow for personas subscribing to sites in their own country. *: Percentage based on news sites which presented cancellation (exit) surveys. Two-step termination (C6) indicates that after pressing cancel, only one further click is needed to terminate subscription. (C7) refers to total amount of clicks required to cancel. }
\label{tab:cancel-data-stats}
\end{table*}

\begin{figure*}
    \centering
    \frame{\includegraphics[width=0.8 \textwidth]{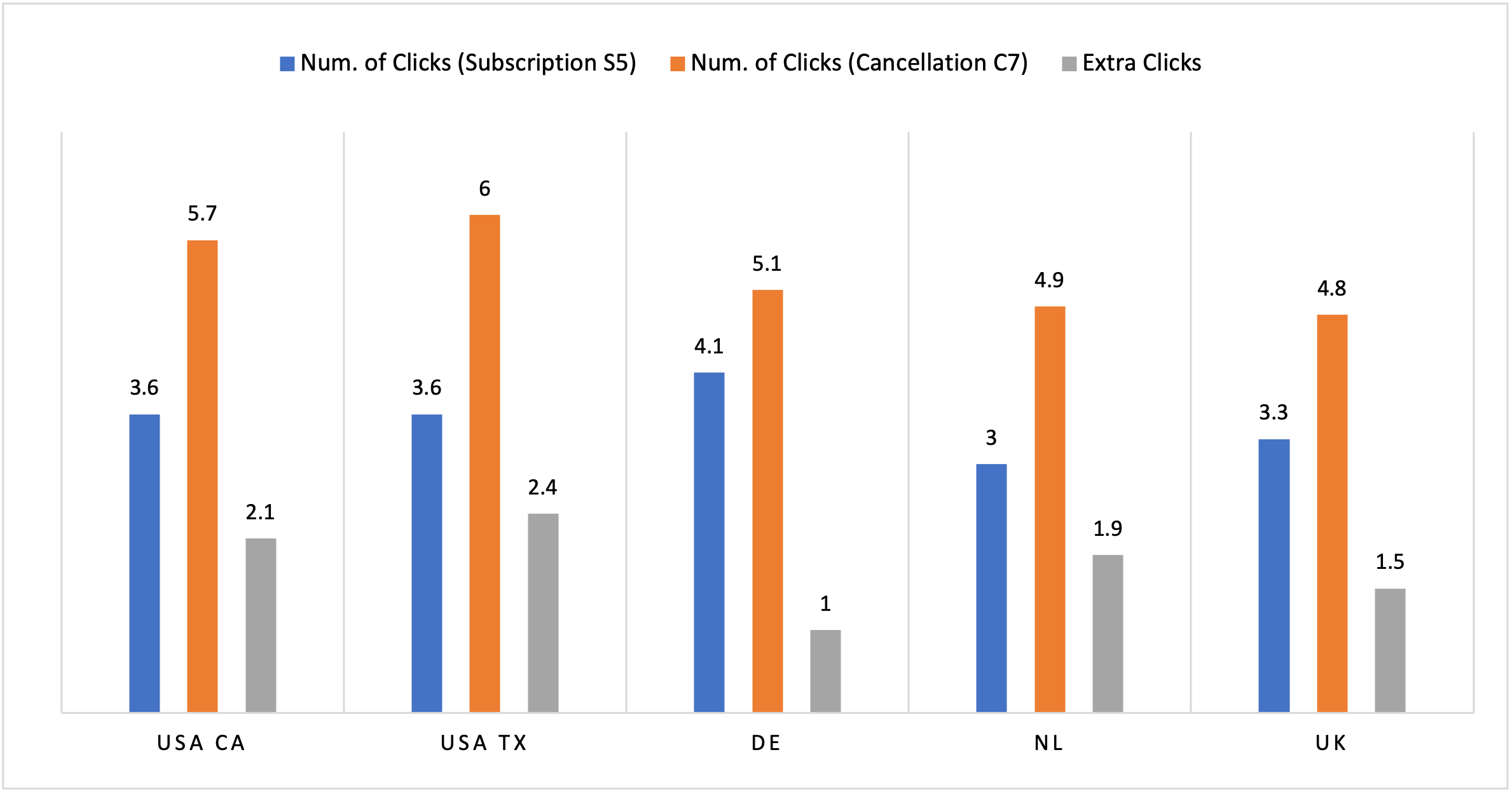}}
    \caption{Average number of clicks it takes for each persona to subscribe and cancel to news sites in their own country. Extra clicks indicate the difference between the number of clicks for cancelling and subscribing. Note: USA CA and TX subsribing to same American news sites, European personas subscribing to national news sites}.
    \label{fig:useuropeclick}
\end{figure*}

The average number of clicks required to subscribe or cancel in each location is given in Figure~\ref{fig:useuropeclick}. European media required fewer clicks for cancellation, compared to US (when subscribed from the US): 4.9 vs 6.2 clicks, respectively.
European media, when accessed from Europe, required on average 3.5 clicks for subscription and 4.9 clicks for cancellation. American sites, again from Europe, averaged 3.2 clicks for subscription and 6.2 clicks for cancellation. Our American personas needed on average 3.5 clicks to subscribe to their American news sites and 5.8 clicks to cancel. 

\begin{figure*}
\centering
\begin{subfigure}{0.4\textwidth}
{\frame{\includegraphics[width=\textwidth]{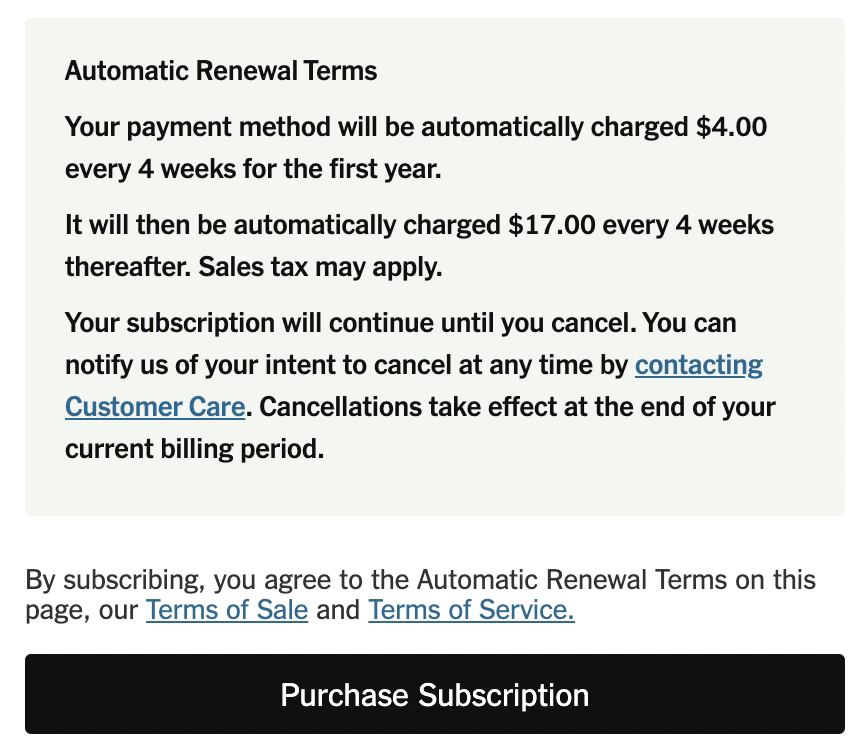}}}    
    \caption{The New York Times}
    \label{fig:NYarNotice}
\end{subfigure}
  \hspace{.5cm}
\begin{subfigure}{0.35\textwidth}
   \frame{ \includegraphics[width=\textwidth]{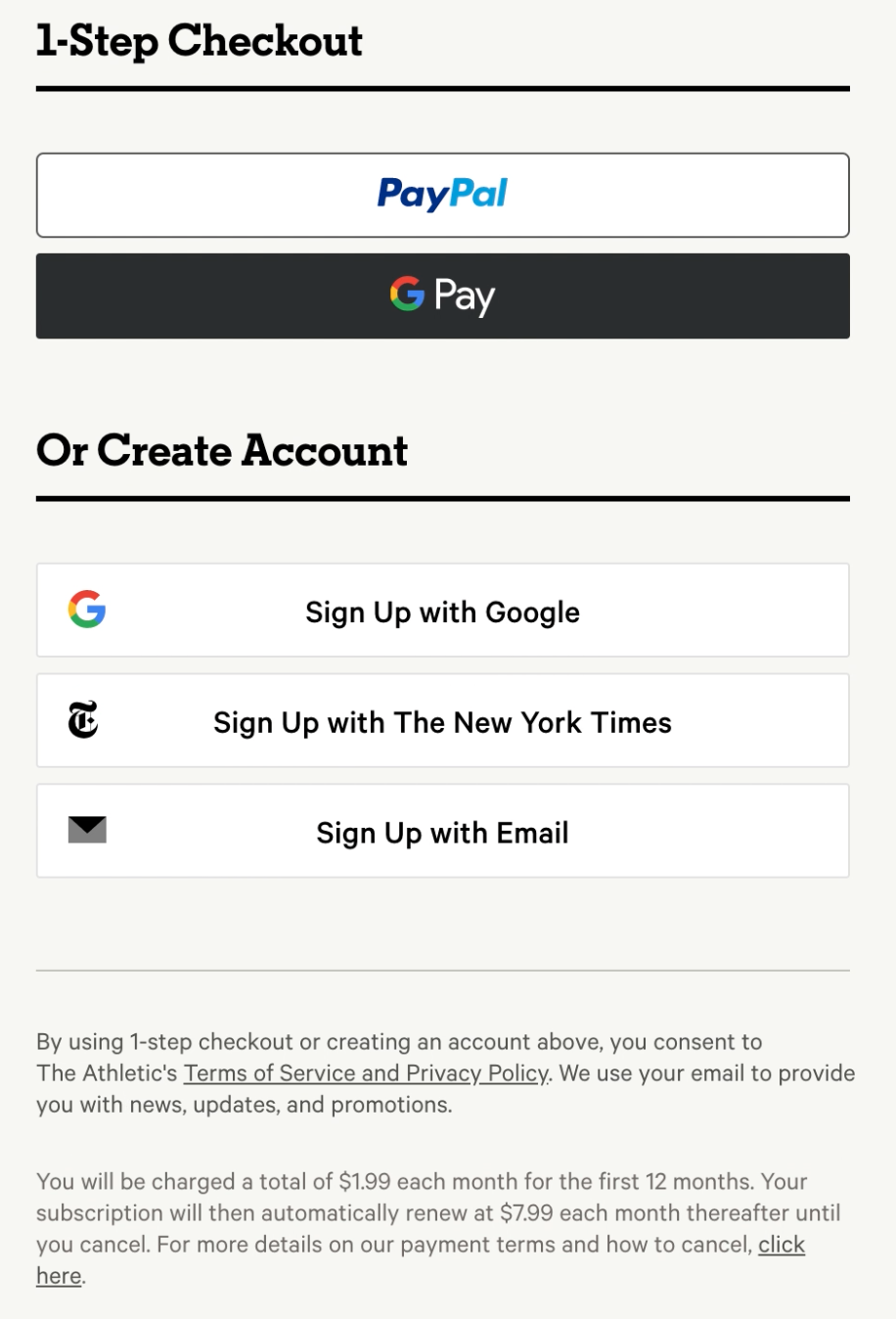}}
    \caption{The Athletic}
    \label{fig:athlectic-small}
  
\end{subfigure}      
\caption{(a) The New York Times' (NYT) auto-renew notice is in bold and notably bigger than the majority of news websites studied. (b) The Athletic's (owned by NYT) checkout page with their renewal information is displayed in small print at the bottom of the dialogue.} 
\label{fig:Ny-Athlectic}
\end{figure*}

\textbf{Auto-renewal disclosure}
All websites except Substack (US) mentioned auto-renewal in their subscription flow\footnote{According to its help page, Substack sends reminders one week before automatic renewal of both monthly and yearly subscriptions~\cite{substack-help-reminders}.
}.
Auto-renewal information was predominantly presented in fine print on the payment page (see Figure~\ref{fig:athlectic-small})\footnote{Screenshots in this paper are taken as part of our data collection in January and February 2023. They are used in this paper in accordance with the \emph{Fair Use}~\cite{fair-use}, \emph{Fair Dealing}~\cite{fair-dealing,fair-dealing-Ireland} and similar doctrines and laws that exempt research related use from copyright restrictions.}. This message informs users that by purchasing a subscription, users acknowledge and agree to the terms of automatic renewal.
Of the 34 individual news sites, 14 auto-renewal messages included a tick box, whereupon ticking indicates that you understand the subscription's auto-renewal terms and conditions. 
Three news sites included an auto-renew button that can be toggled to disable subscription from auto-renewing at the end of the month. This design allows users to pause their subscriptions instead of cancelling them.



\textbf{Information on cancellation}
All European news sites provided information on cancelling subscriptions on the website, either on the homepage or inner pages. In certain cases, this information was found on internal help pages instead of a more prominent location on the page.
In the US, the results were similar: eight out of nine websites provided information about cancellations.

\textbf{Offers to stay \& Exit surveys}
Overall 51.9\% of the European media we subscribed to presented special offers to incentivise staying, and an average of 67.1\% requested reasons for leaving the subscription through the means of an exit survey. Roughly 72.5\% of the European websites made it mandatory to answer these surveys.
For the US roughly 77.8\% of the websites offered discounts to encourage staying, and almost all of the US websites (except Substack, when subscribed from California) displayed an exit survey, with 82.7\% making it mandatory to answer (see Figure~\ref{fig:DW-FORCED-ACTION}).

\textbf{Post-cancellation emails}
After cancelling all subscriptions, 88.9\% of the American and 92.1\% of the European news websites sent an email confirming cancellation. Exceptions to this were the Noordhollands Dagblad (NL) and the Guardian (UK).
Once subscriptions were cancelled for all accounts, our personas continued to receive emails asking them to re-subscribe and providing news updates.

Below, we discuss the results in more detail for each country and we give an overview of dark patterns encountered during the subscription and cancellation processes.

\subsection{United States of America}
\label{sec:USAresults}
\subsubsection{Subscribing}
We successfully subscribed to nine of the ten American news sites using our Texan and Californian personas.
All American and European personas failed to subscribe to the LA Times, possibly due to the use of our VPN. 
Our European personas were additonally blocked by Barron's due to location, and subscribed to eight of the ten American news sites.

Eight of the nine news sites informed our personas about automatic renewal, Substack provided auto-renewal information only on their help centre page. Substack also provided different suggested payments (or pledges) for subscribing ranging from \$5 a month to \$150 a year or no pledge. 
Four of the nine American websites subscribed to from our Californian persona featured an auto-renew acknowledgement tick box. For our Texan consumer, Barron's and Wall Street Journal did not feature this tick box, which was available to our Californian consumer. The remaining websites informed consumers of their renewal policy with a notification in the payment form in small print.
Despite not having a tick box, The New York Times' notice was clearer and easier to read (Figure~\ref{fig:NYarNotice}). 


Six of the nine American news sites provided information on cancelling during the subscription flow, and 
all websites provided a post-subscription email in which they notified consumers of their renewal policy.
The Washington Post, Medium and The Daily Wire were the only websites that clearly stated the next billing date in their post-subscription email. After subscribing to the Daily Wire's monthly subscription, we received several promotional emails inviting us to upgrade to their annual plan.

Comparing the number of clicks required to subscribe, we find similar results for our American and European personas subscribing to American newspapers. On average, our American personas made 3.5 clicks, while our European personas made 3.2 clicks when subscribing to American websites (see Table~\ref{tab:clicks} in Appendix~\ref{app:C}).

\subsubsection{Cancelling}
It was possible to cancel all Californian subscriptions online, however cancelling the Wall Street Journal for our Texan and British consumers required making a phone call. The calls were no more than three-minutes long. The customer representative requested our email address, and our address or the last few digits on the credit card. 
They also asked the reason for terminating the subscriptions, but did not offer discounts or special offers during the calls.
Shortly following the calls, a cancellation confirmation email was received. 

\begin{figure*}
\centering
\begin{subfigure}{0.3\textwidth}
{\frame{\includegraphics[width=\textwidth]{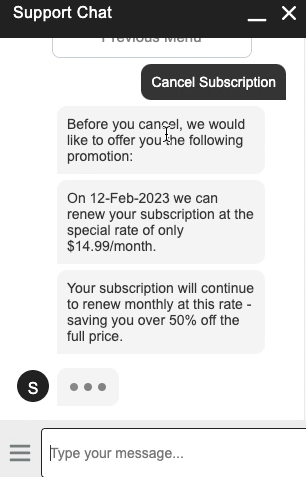}}}    
    \caption{Bloomberg response after choosing to cancel from a list of options.}
    \label{fig:BloombergChat1}
\end{subfigure}
  \hspace{.5cm}
\begin{subfigure}{0.3\textwidth}
   \frame{ \includegraphics[width=\textwidth]{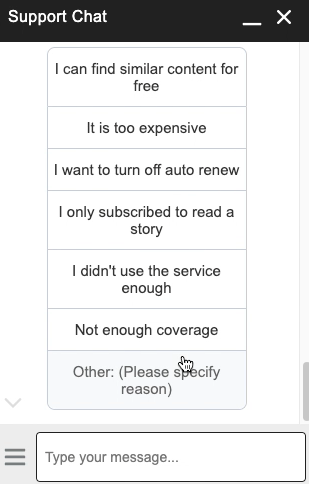}}
    \caption{After cancelling, Bloomberg requesting you answer a survey.}
    \label{fig:BloombergChat2}
\end{subfigure}      
\caption{Bloomberg's chatbot cancelling procedure.} 
\label{fig:BloombergChat}
\end{figure*}

Barron's and the Wall Street Journal are both published by Dow Jones \& Company. As a result, when purchasing both subscriptions, they appear together on either website's settings. Despite this, it is possible to cancel Barron's online both in Texas and California, whereas Wall Street Journal requires calling to cancel from Texas. The Bloomberg subscription required cancellation via a closed-domain~\cite{closed-domain-chatbot} chatbot for all personas, which required a total of ten clicks to cancel. 
The chatbot offered promotional discounts and requested reasons for leaving (Figure~\ref{fig:BloombergChat}).

With the exception of Substack for California, all American news sites requested reasons for cancelling the subscription. Of these, 77.7\% made answering mandatory. 
Seven and six of the nine American news sites offered discounts to stay to our Texan and Californian personas, respectively. When cancelling Barron's from Texas, discount offers were presented, but not when cancelling from California. For the UK and Germany, Substack offered discounts.
All media apart from Bloomberg sent a confirmation of cancellation email. Nineteen days after cancelling The Daily Wire subscription, all personas received a promotional email with an offer to re-subscribe. The initial email was followed by another email two days later, reiterating the offer with the line \say{No pressure just wanted to bump this promotion}. Yet again nine days later, another email is received notifying all personas that the offer is expiring. The Daily Wire was the only website that sent third-party promotions via email. Some of these emails requested donations for causes such as anti-abortion campaigns and the Prison Fellowship who supply Bibles to prisons. 

\begin{figure*}
    \centering
    \frame{\includegraphics[width=\textwidth]{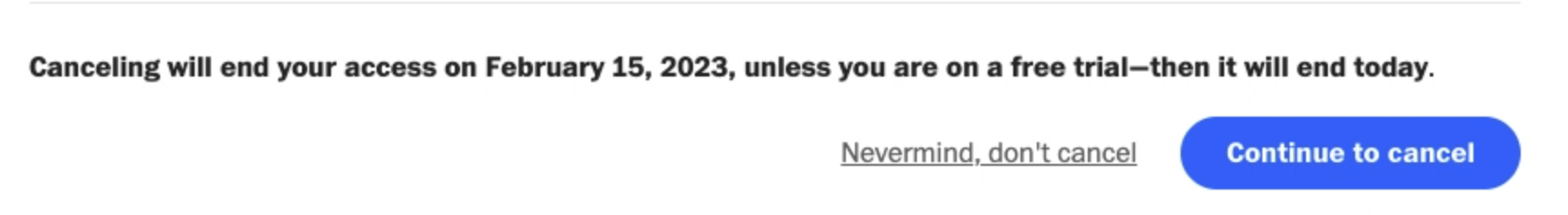}}
    \caption{An example of a \textit{bright pattern} captured in our experiments: The Washington Post's cancellation interface presents the ``Continue to cancel'' button more prominently, contrary to the majority of media cancellation interfaces.}
    \label{fig:WaPoBright}
\end{figure*}

The Washington Post's cancellation flow stood out from others, because its design reaffirmed users' intent to cancel by presenting the `Continue to cancel' button more prominently (Figure~\ref{fig:WaPoBright}).
This is the opposite of traditional Visual Interference dark pattern that presents the cancellation button less prominently to dissuade users from leaving the service.

\subsection{Germany}
\subsubsection{Subscribing}
Our German persona was successful in subscribing to eight German media websites. Rheinpfalz Plus and Allgemeine Zeitung+ required real German addresses for a subscription, which we could not provide. 

\begin{figure*}
    \centering
    \frame{\includegraphics[width=\textwidth]{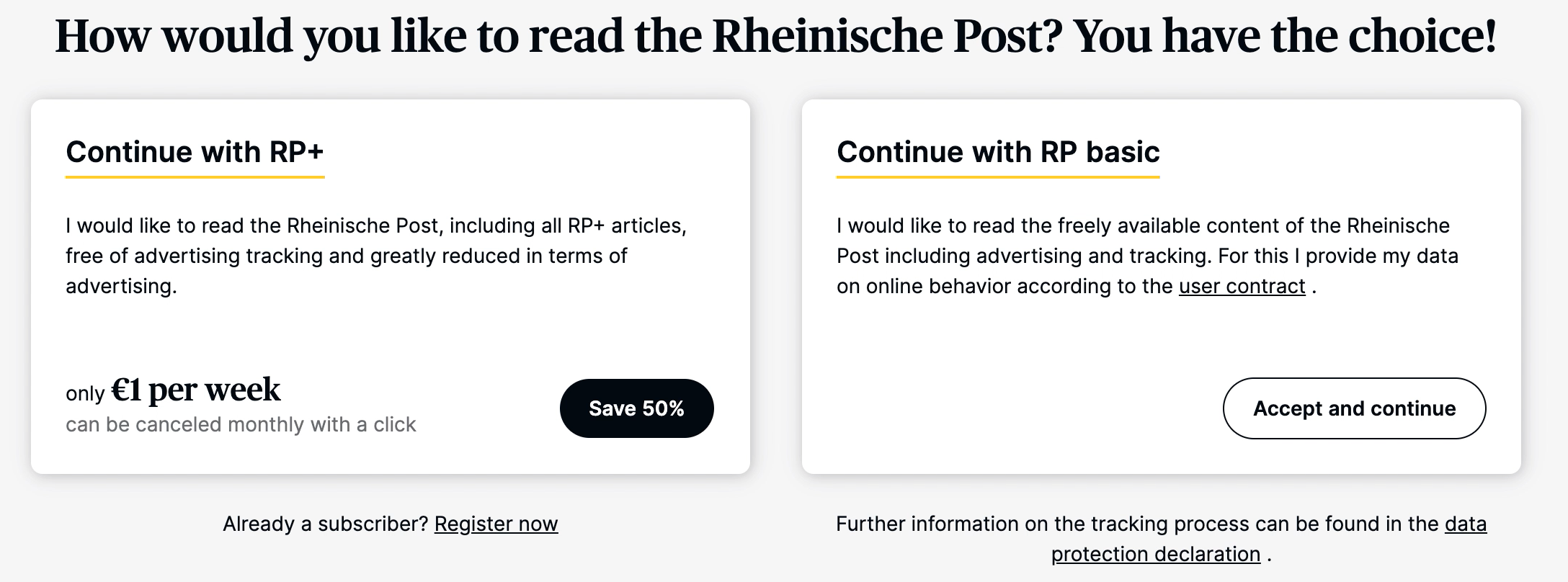}}
    \caption{The  \textit{cookie paywall}~\cite{morel2022your} on Rheinische Post's subscription page. The user is asked to choose between (left) paying for a subscription with reduced ads and tracking, and (right) free access with more ads and tracking.} 
    \label{fig:RP-CookieWall}
\end{figure*}

Five of the news websites displayed a \textit{cookie paywall} (see Fig~\ref{fig:RP-CookieWall}), where the user is presented with a paid subscription option that limits tracking and advertisements. 
In our study, when presented with a cookie paywall, we always opted for the paid subscription.
In their 2022 study, \citet{morel2022your} found that cookie paywalls are mostly used in news sites and do not prevent all tracking. In Europe at present, there is a lack of consensus among data protection agencies (DPAs) as to whether cookie paywalls comply with GDPR requirements ~\cite{morel2022your}.

All websites provided notice of automatic renewal, and information on cancelling was provided during the subscription flow. The only exception to this was Heise, who provided no information on cancelling in their subscription flow.

All websites sent confirmation emails after subscribing. Kieler Nachrichten+ was the only news website to state a renewal date in their post-subscription email. Bildplus and the Pioneer attached receipts with account information in their emails.

\subsubsection{Cancelling}

It was possible to cancel all German media subscriptions online. Four out of the eight websites offered discounts to stay.
The Pioneer required filling in a form to cancel the subscription. This form requested full name, email and invoice number, the last of which was not mandatory. 
Out of the eight German subscriptions, four inquired as to the reason for cancellation through the means of a survey, three of which were mandatory to answer. 
All websites sent emails confirming the cancellation; some of these contained further offers.  
KN+ was the only German website that featured an auto-renew button, which could be switched off.

\begin{figure*}
  \centering 
  \frame{\subfloat[]{\includegraphics[width=0.445\textwidth]{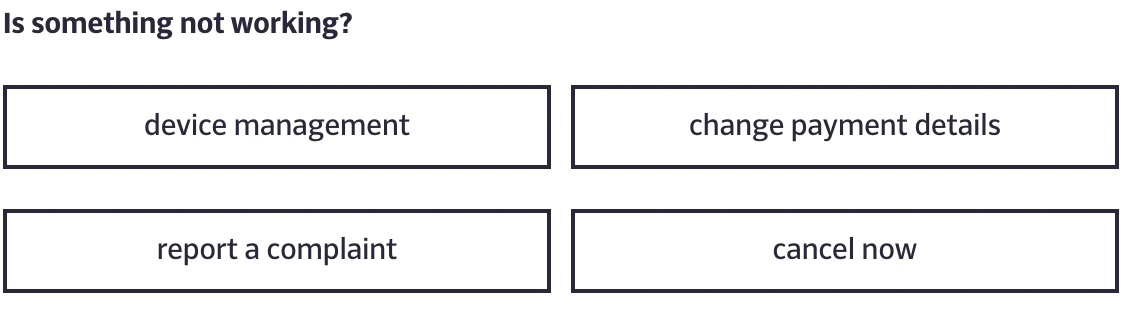}}} 
  \frame{\subfloat[]{\includegraphics[width=0.212\textwidth]{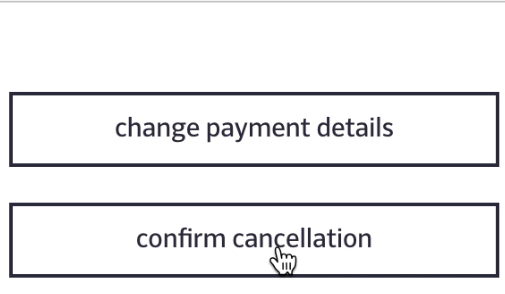}} }
  \frame{\subfloat[]{\includegraphics[width=0.21\textwidth]{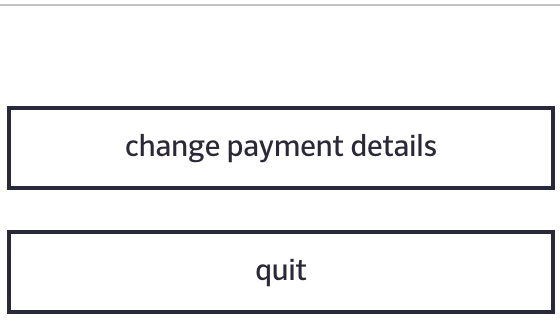}}}
  \caption{Süddeutsche Zeitung+'s two-click termination button.}
  \label{fig:SZCancel}
\end{figure*}

The Fair Consumer Contracts Act (FCCA) states it should take no more than one extra step after pressing the `Cancel now' button to successfully cancel your subscription \cite{cancelbutton}. In our study of German websites, we measured the number of clicks required to cancel once the `cancel now' button was pressed. 
Only three of the eight websites satisfied the FCCA's requirement. 
Figure~\ref{fig:SZCancel} shows an example of cancelling a subscription in two clicks, part (a) shows a clear `cancel now' button. After clicking this, the button changes to a `confirm cancellation' button (b). The subscription is cancelled when the `confirm cancellation' button is clicked. 

\begin{figure}
        \centering
        \frame{\includegraphics[width=0.50\textwidth]{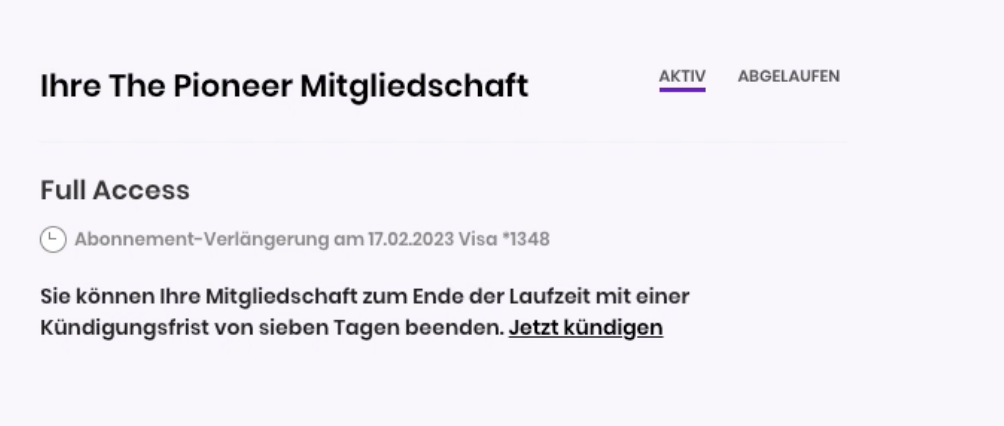}}
    \caption{The link for cancelling Pioneer's subscription is underlined (``Jetz kündigen''), but it is much less prominent than a button---which the FCCA requires.} 
    
       \label{fig:subtlelinks}
\end{figure}

The FCCA also stipulates that the `termination button' must be prominently displayed and appropriately labelled. Certain links encountered on German news sites were not clear or obvious, compared to a clearly visible button. For instance, while Pioneer's cancel link in Figure~\ref{fig:subtlelinks} is underlined, it is unclear whether it constitutes a prominently displayed \textit{button}. 

\subsection{The Netherlands}
\subsubsection{Subscribing}
Our Dutch persona was successful in subscribing to nine of the ten Dutch news sites on our list. Het Financieele Dagblad (FD) refused to accept our (Revolut) account number. All Dutch websites required direct debit payments and sent post-subscription emails, none of which mentioned billing dates.  
All websites provided information on automatic renewal and cancelling, both in the subscription flow and the website interface. All websites also provided a tick box to acknowledge auto-renewal or terms and conditions. 

\subsubsection{Cancelling}
\begin{figure*}
   \centering
    \begin{subfigure}[b]{0.44\textwidth}
        \centering
        \frame{\includegraphics[width=\textwidth]{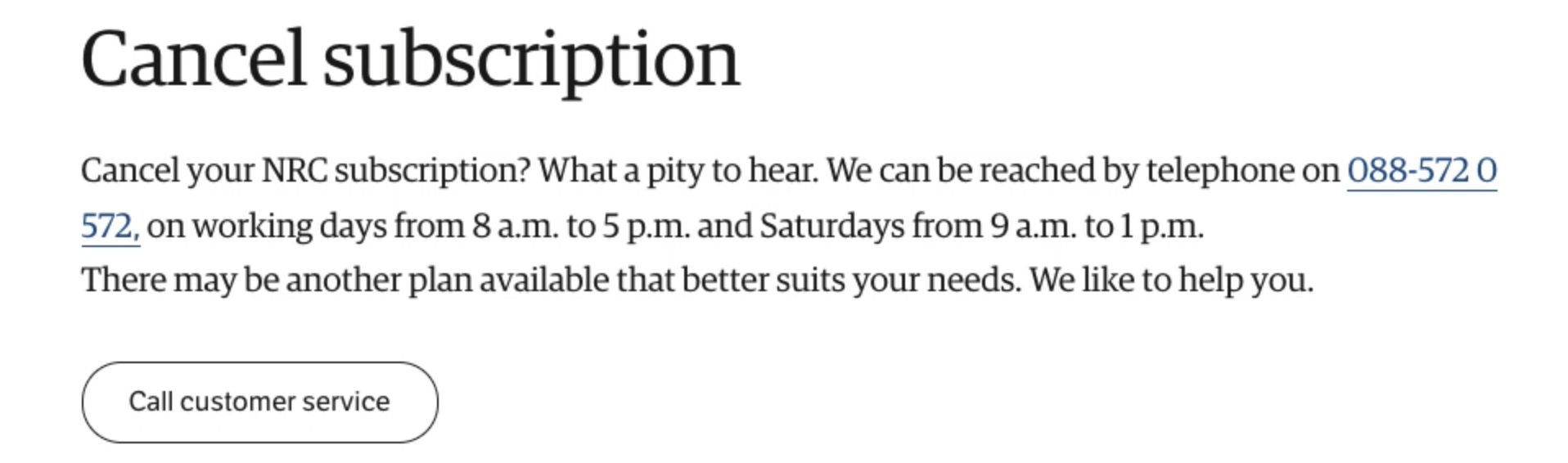}}
        \caption{NRC}
        \label{fig:NRC-call-to-cancel}
    \end{subfigure}
   \vspace{10pt} 
    \begin{subfigure}[b]{0.485\textwidth}
        \centering
        \frame{\includegraphics[width=\textwidth]{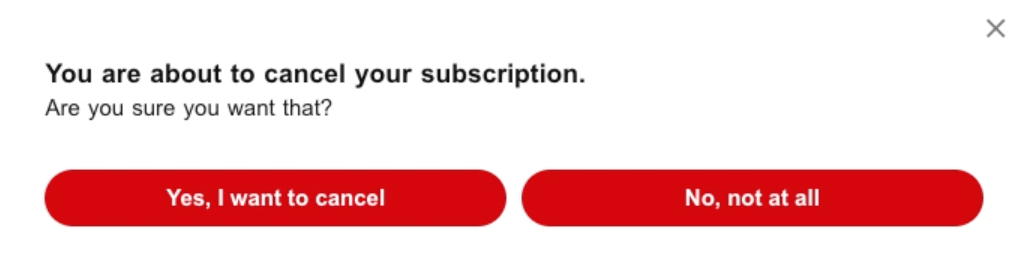}}
        \caption{Het Algemeen Dagblad (AD)}
        \label{fig:ADcolourcancel}
    \end{subfigure}
    \caption{(a) NRC is the only Dutch newspaper that requires a phone call for cancelling subscriptions. 
    (b) Het Algemeen Dagblad (AD) cancel notice with equal colours. }
       
\end{figure*}

It was possible to cancel all subscriptions online except for the public broadcaster NRC. Cancelling NRC required a phone call (see Figure\ref{fig:NRC-call-to-cancel}), during which we were on hold for ten minutes. 
When cancelling our Dutch media sites, all sites apart from NRC requested a reason for leaving. De Telegraaf and Noordhollands Dagblad required us to fill in an online form that included an optional survey about reasons for leaving. We received cancellation emails for all subscriptions bar one (Noordhollands Dagblad). 
On average, 55.6\% of Dutch news sites presented special offers to stay, but notably, these offers were often positioned on the same page as the `cancel now' button. This contrasts with practices in other countries where such offers typically appear after starting the cancellation process. 
Most Dutch websites required a confirmation after clicking cancel, and the options in the confirmation dialogue were presented equally (see Figure\ref{fig:ADcolourcancel}).

\subsection{United Kingdom}
\subsubsection{Subscribing}
Our British persona successfully subscribed to eight out of the ten British websites. The Times and The Week news sites rejected our virtual credit card. All websites provided information on cancelling and automatic renewal.
All websites sent a confirmation of subscription email, which varied in the amount of details given regarding the subscription. For example, The Daily Mail, The Guardian and Tortoise provided no details about renewal dates. The Guardian presented different prices ranging from £3 to £10 as contribution which unlocks different features. We chose to pay £3.

\subsubsection{Cancelling}
\begin{figure}
    \centering
    \frame{\includegraphics[width=0.5 \textwidth]{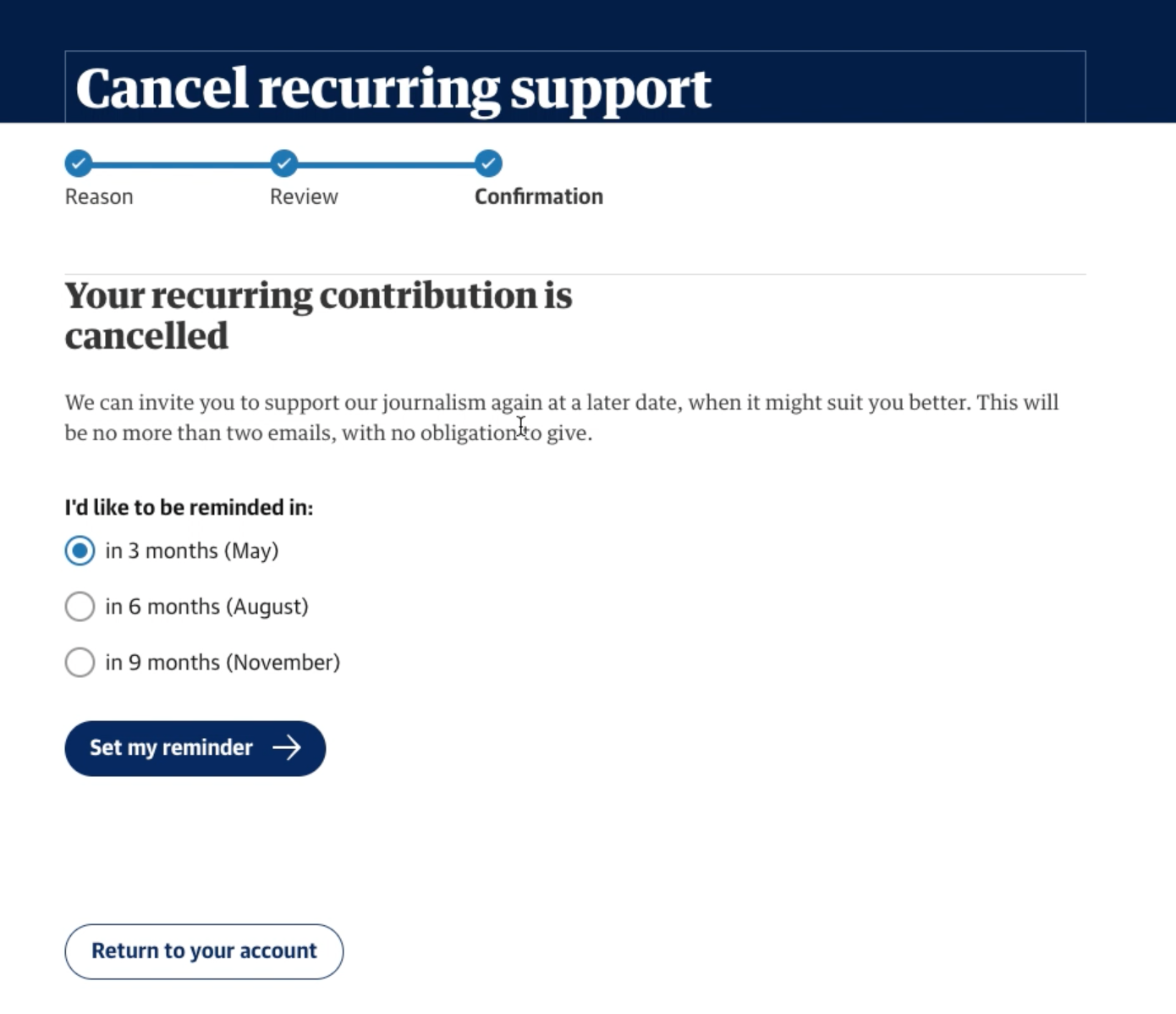}}
    \caption{An example of a \emph{bright pattern} from The Guardian, where they ask if you would like to be reminded to resubscribe at a different stage with choices of 3, 6 or 9 months later. }
    \label{fig:guardian-bright-pattern}
\end{figure}

With the exception of The Daily Mail, it was possible to cancel all accounts online.
The Spectator required us to fill in a Google Form embedded on their website. This form requested full name and email which was mandatory. The user's address and subscriber number were also requested, but were not mandatory.
Tortoise and iNews both had an auto-renew button that could be toggled to disable auto-renewal without cancelling the subscription.
Out of the eight British websites our British persona subscribed to, five requested a reason for cancelling in the form of a survey, and four of these websites made it obligatory to provide an answer. Four out of the eight websites offered discounts before confirming cancellation. 
All news sites apart from The Guardian provided confirmation emails after cancelling accounts.
The Guardian presented a unique feature among all studied news sites. After the cancellation process, it offered the option to receive email reminders after 3, 6, or 9 months, encouraging to consider contributing once again (refer to Figure~\ref{fig:guardian-bright-pattern}).


%% file: sections/dark-patterns.tex
\subsection{Dark Patterns}
\label{subsec:DP}

In this section we present examples of dark patterns observed in the graphical user interface (GUI) and interaction flow.
As many dark patterns were repeated in similar ways in all countries, we present them in this separate section to avoid repetition. We do not claim to present an exhaustive list of dark patterns, which is out of scope for our study. Instead, we aim to share representative examples based on the taxonomy of \citet{mathur2019dark}. We focus on \emph{Hard to cancel}, \emph{Forced Action}, \emph{Visual Interference}, and \emph{Confirmshaming} dark patterns.
These patterns were mostly found in cancellation flows, apart from \emph{Forced Action}, which were also present in subscription flows.

\textbf{Hard to cancel:}
The most notable barrier to cancelling was having to click through numerous steps of (often mandatory) surveys,  promotional offers and confirmation dialogues before finally cancelling the service. For instance, unsubscribing from Bloomberg required ten clicks, and it involved launching and interacting with a closed-domain chatbot (Figure~\ref{fig:BloombergChat}). 
In particular, in order to cancel, a Bloomberg user needed to click \textit{Your account} $\rightarrow$ \textit{Subscription} $\rightarrow$ \textit{Contact Support to Cancel} $\rightarrow$ \textit{(Read the help article)} $\rightarrow$ \textit{Launch Chat} $\rightarrow$ \textit{(Enter first and last name on the chatbot welcome page)} $\rightarrow$ \textit{Start Chatting} $\rightarrow$ \textit{Manage my Subscription} $\rightarrow$ \textit{Cancel Subscription} $\rightarrow$ \textit{No} (\textit{to discount offers}) $\rightarrow$ \textit{Yes} (\textit{to question to confirm cancellation}) $\rightarrow$ \textit{(Answer the mandatory exit survey)} $\rightarrow$ (Cancellation request received).

In other hard to cancel subscriptions,
multiple low-level design patterns were combined to have an aggregate `Roach Motel' effect~\cite{Brignull}. We study such patterns in the following.

\begin{figure*}
     \centering
    \begin{subfigure}[b]{0.6\textwidth}
         \centering
         \frame{\includegraphics[width=\textwidth]{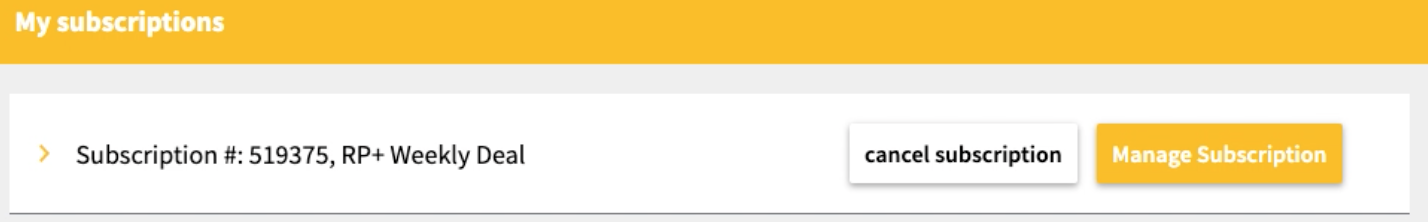}}
         \caption{`Cancel subscription' button is white in the first dialogue}
     \end{subfigure}
     \begin{subfigure}[b]{0.85\textwidth}
         \centering
         \frame{\includegraphics[width=0.7\textwidth]{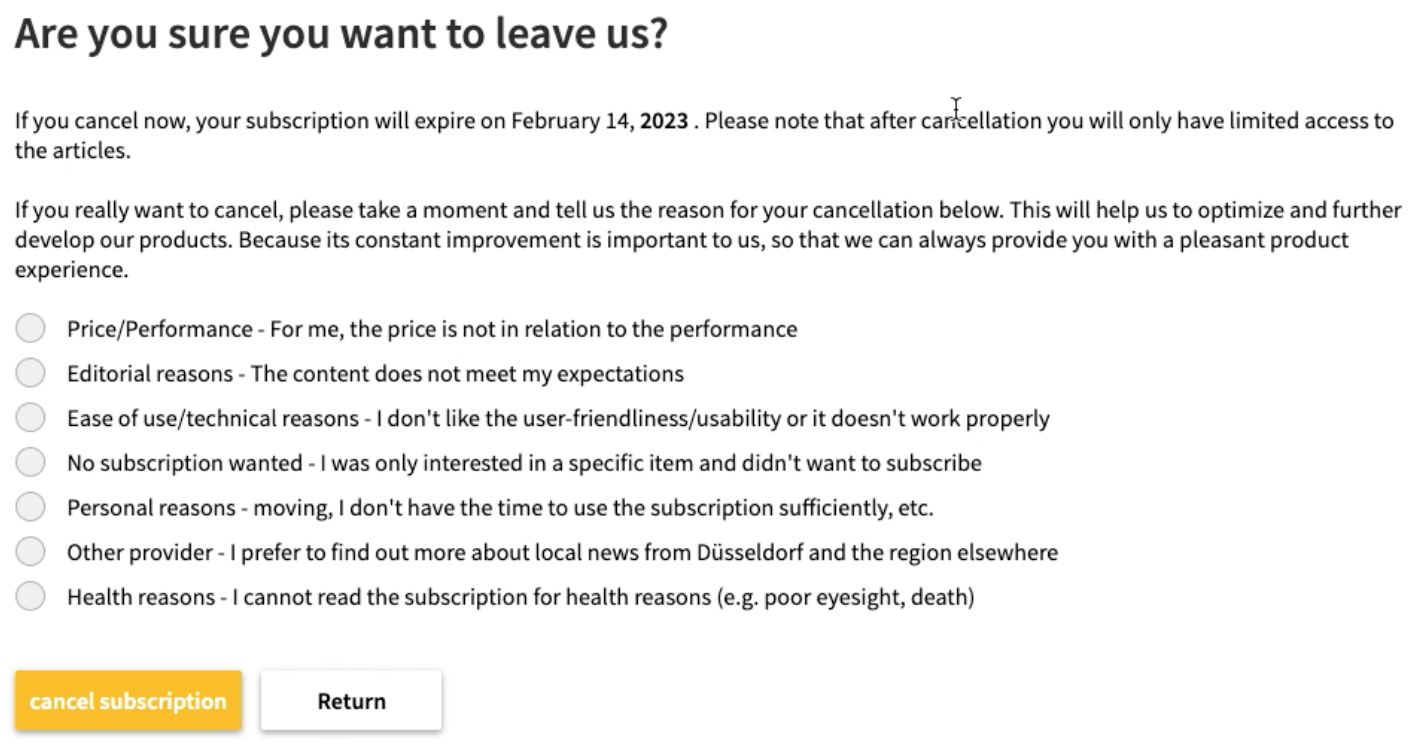}}
         \caption{`Cancel subscription' button is yellow in the second dialogue. Clicking again the white (`Return') button would terminate the cancellation process.}
     \end{subfigure}
     \caption{Rheinische Post's cancellation flow cancel button featuring visual interference.}
        \label{fig:RP+ColourSwitch}
 \end{figure*}

\textbf{Visual Interference}
\citet{mathur2019dark} characterizes \emph{Visual Interference} as use of style and visual presentation to influence users' choices.
We found many cases of alternating colours between the `cancel' button and the `keep subscription' button, which is also categorized as \textit{Bait and switch} by Brignull~\cite{Brignull}.
An example of this can be seen can be seen in Figure~\ref{fig:RP+ColourSwitch} from the German newspaper Rheinische Post. The `Cancel subscription' button is initially white and the `Manage subscription' button is yellow. After pressing `Cancel subscription' however, the next page now displays the `Cancel subscription' button in yellow and the `Return' button in white. This form of misdirection \cite{mathur2019dark} can potentially lead a consumer to inadvertently exit the cancellation flow by pressing the wrong button.

\begin{figure*}
    \centering
    \frame{\includegraphics[width= 0.6\textwidth]{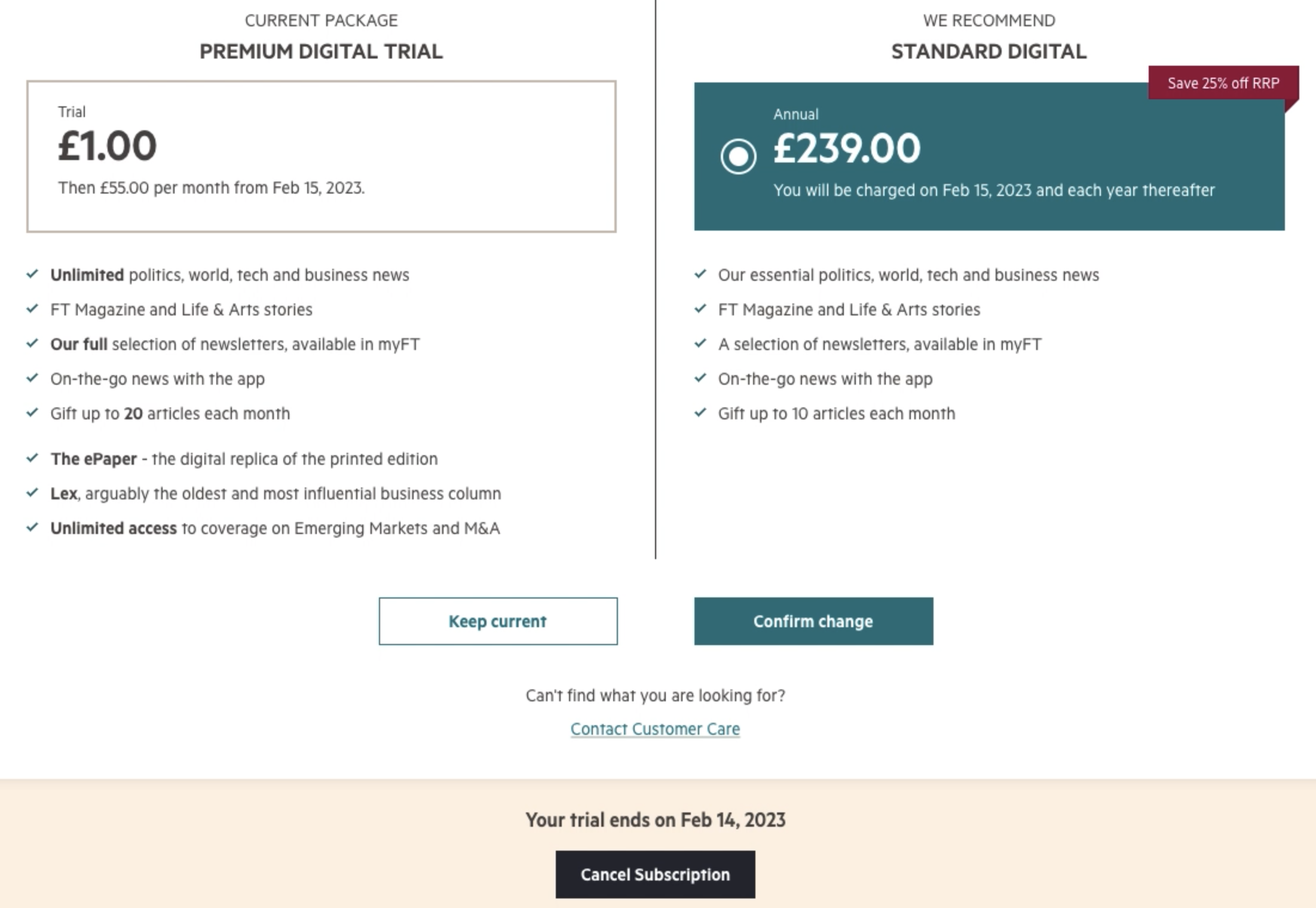}}
    \caption{\textit{Visual Interference example} from the Financial Times: the `Confirm change' button confirms purchasing an annual offer; it does not confirm the cancellation---which is the intended action of the user. The actual `Cancel subscription' button placed at the bottom of the page and it can be easily missed.}
    \label{fig:FToffer}
\end{figure*}

\begin{figure*}
    \centering
    \frame{\includegraphics[width=0.6 \textwidth]{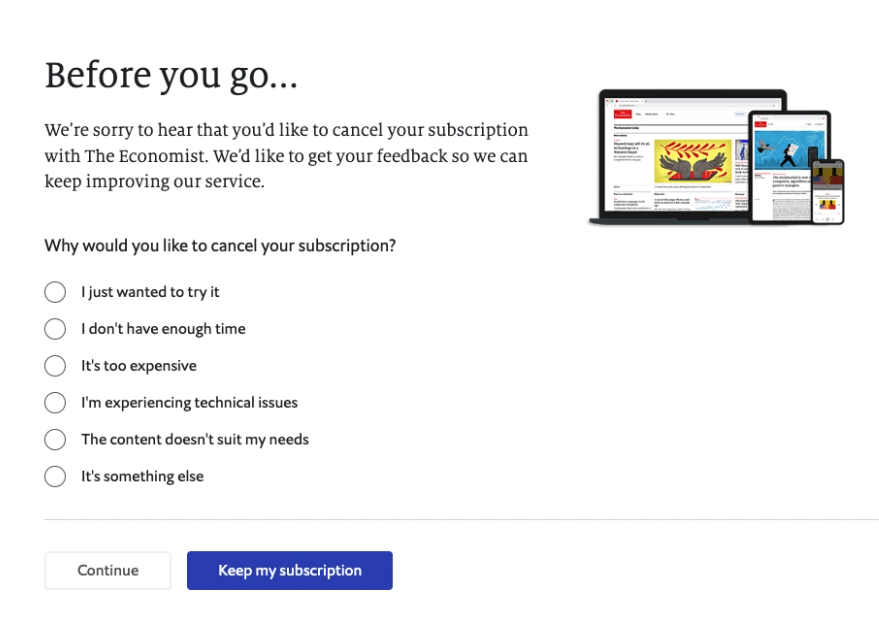}}
    \caption{\emph{Visual Interference} example from The Economist. Here on their exit survey page, the `continue' cancellation button is in white and the `keep subscription' button is in a more visible colour blue. }
    \label{fig:Economist-VI}
\end{figure*}

Another example of \emph{Visual Interference} is misleading buttons in a promotional offer from the Financial Times when cancelling. After starting the cancellation process, our British persona is presented with her current subscription and a yearly subscription---the latter of which is unexpectedly chosen by default.
Clicking the `Confirm change' on this interface upgrades our subscription to a yearly one, rather than cancelling it. The actual `Confirm cancel' button is placed at the bottom of the page, which can be easily missed (see Figure~\ref{fig:FToffer}). Finally, The Economist  displays their `Continue' with cancellation button in white (same as the background) and their `Keep subscription' button in blue, which is visually more prominent (see Figure~\ref{fig:Economist-VI}). 

Although the majority of websites provided information on auto-renew, oftentimes the information was in small print and could be overlooked. Cancel buttons were often hard to locate, hidden amongst plain text (Figure~\ref{fig:subtlelinks}) or placed at the bottom of the dialogue where they could be missed (Figure~\ref{fig:FToffer}).

\begin{figure}
\centering
\begin{subfigure}{0.45\textwidth}
    \frame{\includegraphics[width=\textwidth]{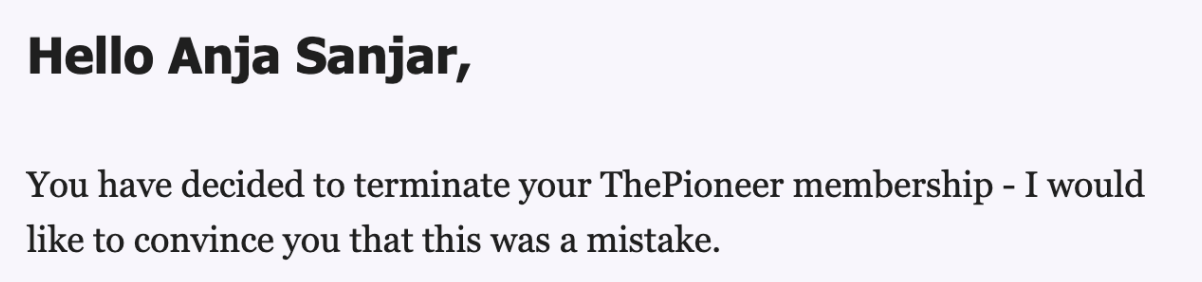}}    
    \caption{The Pioneer}
\end{subfigure}
  \hspace{.5cm}
\begin{subfigure}{0.45\textwidth}
   \frame{ \includegraphics[width=\textwidth]{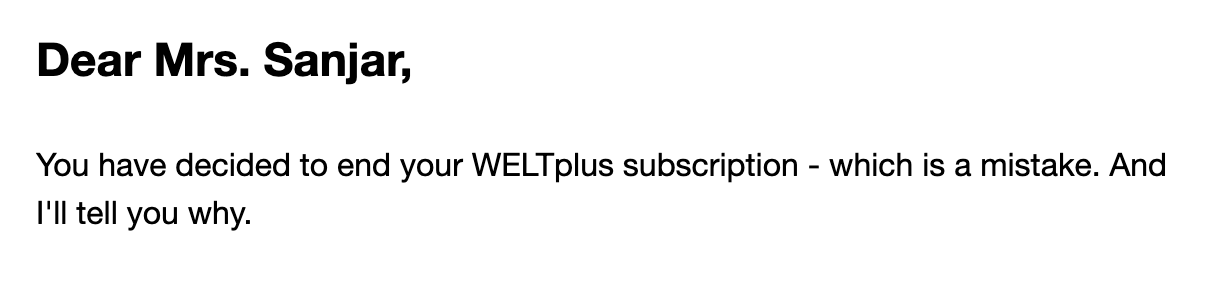}}
    \caption{Die Welt}
\end{subfigure}      
\caption{Messages received after cancelling subscriptions from The Pioneer's news sites and Die Welt. }
\label{fig:darkemails}
\end{figure}

\textbf{Confirmshaming}
We found several examples of \emph{Confirmshaming} dark patterns in post-cancellation emails. 
The Pioneer's post-cancellation email stated \say{Journalism is a participatory event} and they would like us to \say{be a part of the party again}. In certain cases, the tone approached toward accusatory, calling our decision to cancel our subcription a `mistake' (see Figure~\ref{fig:darkemails}).
In particular, The Pioneer's email said ``Hello Anja Sanjar, you have decided to terminate your The Pioneer subscription - I would like to convince you that this was a mistake''. The email from Weltplus said ``you have decided to end your Weltplus subscription-which is a mistake. And I'll tell you why".
In addition to post-cancellation emails, dialogues within the cancellation flow contained confirmshaming patterns. For instance, immediately after finishing the cancellation De Volkskrant showed a pop-up dialogue with the message: \say{What a pity you are leaving!}.

\begin{figure*}
\centering
\begin{subfigure}{0.44\textwidth}
{\frame{\includegraphics[width=\textwidth]{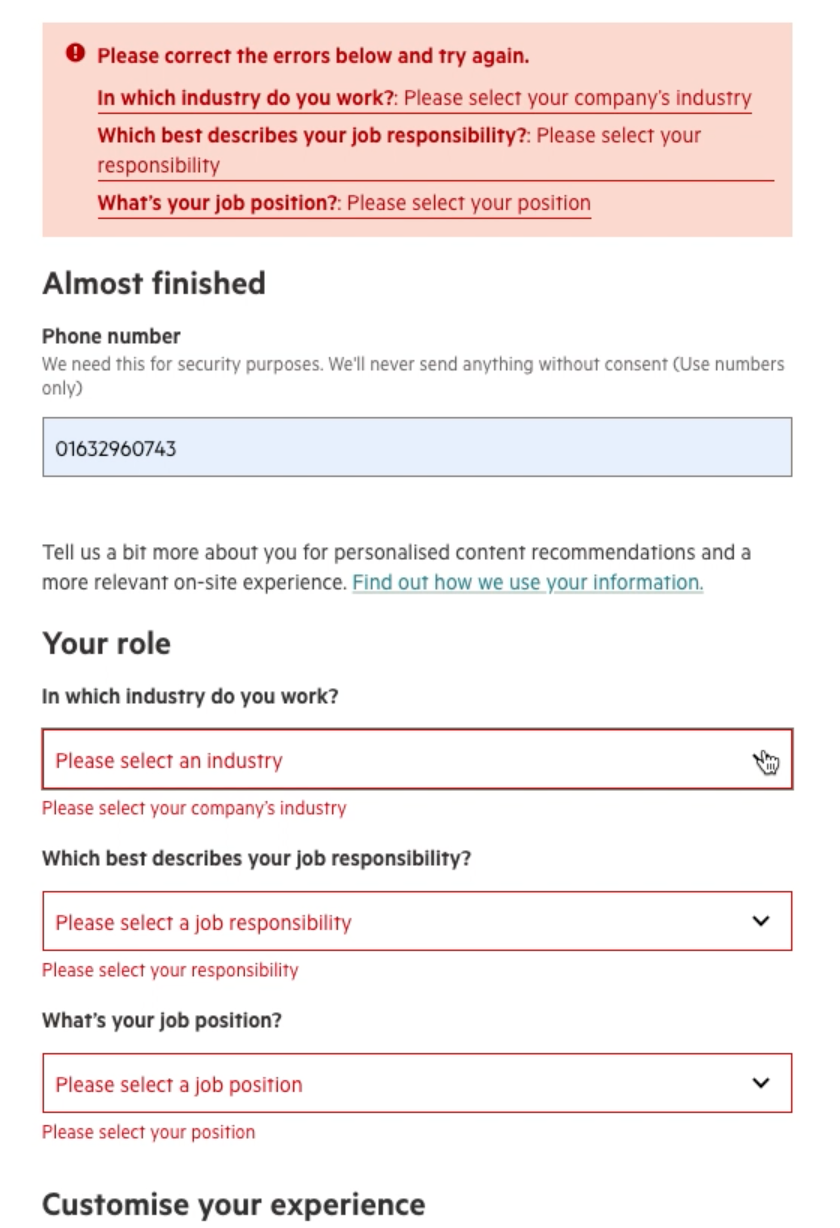}}}    
    \caption{Financial Times}
    \label{fig:FT-mandatory}
\end{subfigure}
  \hspace{.5cm}
\begin{subfigure}{0.475\textwidth}
   \frame{ \includegraphics[width=\textwidth]{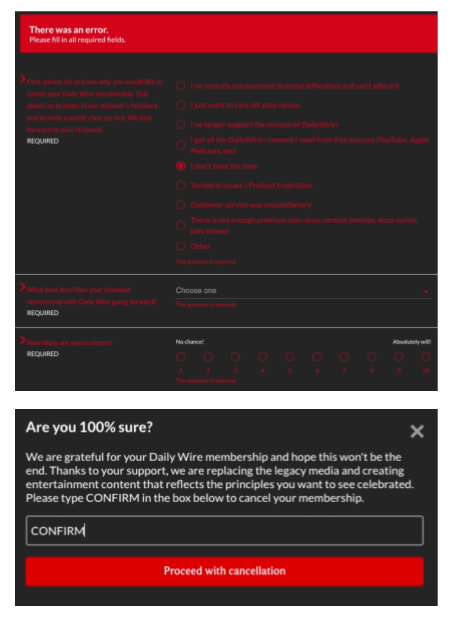}}
    \caption{Daily Wire}
    \label{fig:DW-FORCED-ACTION}
  
\end{subfigure}      
\caption{Examples of \textit{Forced Action}. (a) Mandatory information required to sign up for the Financial Times subscription. 
(b) The Daily Wire makes it mandatory to answer survey questions during cancellation and type the word `CONFIRM' to cancel.} 
\label{fig:FT&DWmand}
\end{figure*}

\textbf{Forced Action}
Designs that coerce users to take a generally unnecessary action to complete a task are categorized as 
\emph{Forced Action}.
The Daily Wire, for instance, required users to answer a 3-question mandatory survey and
type in `CONFIRM' to complete cancellation (see Figure~\ref{fig:DW-FORCED-ACTION}).
Some amount of friction (e.g., a confirmation dialogue) may help prevent accidental cancellations, however, requiring users to type a phrase appears unnecessary. This pattern is unique to The Daily Wire in our complete dataset.

Turning to subscriptions, while requiring users to create an account is expected and understandable, certain newspapers required more than necessary information during subscription (Table~\ref{tab:sub-data-summary}). \citet{bosch2016tales} refers to excessive data collection as a \emph{Maximise} privacy dark strategy. 
Overall, the type of data collected during subscription varied across locations (see Table~\ref{tab:sub-data-summary}) . 
The American and British news websites required  more user data during sign-ups than their Dutch and German counterparts. Several websites asked for extensive personal details, including full address, date of birth, occupation, and phone number. 
In certain instances, providing this information was mandatory, leading to a cumbersome registration process (see Figure~\ref{fig:FT-mandatory}).

In summary, while careful or tech-savvy users may navigate these design tricks to complete their cancellations, users with low tech literacy, reduced vision or confused mental state may easily fall for these traps and end up paying for subscriptions that they do not need and use.

%% file: sections/discussion.tex
\section{Discussion}
\label{sec:discussion}


Certain regulations require cancellation to be as easy as subscribing. 
There is risk however, that these long and complex subscription flows will be misused to bypass such regulation.
For example, the combination of dark patterns in the graphical user interface and interaction flow can be effective at encouraging users to avoid cancelling in a more nuanced way compared to an obvious deterrent, such as forcing a user to phone to cancel. \citet{gray2018dark} highlight the ethical concerns around persuasive design.
When cancelling, websites often took us through many different steps (up to ten) before reaching the end goal of exiting the contract. These obstacles included locating the cancellation button or link, answering mandatory survey questions, and (in one case) typing in a specific phrase.
While subscribing was usually less circuitous, 
certain subscriptions required a considerable amount of personal information disclosure (Table~\ref{tab:sub-data-summary}).

Cancelling online was possible for 63 of the 67 subscriptions made in the study, but these were not effortless processes.
Four of the 67 subscriptions could only be terminated by contacting the customer representative via phone. Calling to cancel, in most cases, was straightforward, and often the phone number was easier to locate than an unsubscribe button. 
Out of the four phone calls placed, three lasted approximately three minutes each.
The remaining call (NRC, NL), however, took more than ten minutes. As mentioned in Section~\ref{sec:intro}, requiring users to make phone calls is often associated with cognitive burdens insofar as anticipating and engaging in phone conversations can evoke anxiety in certain individuals. While offering the choice to cancel via phone can be beneficial, it may not be a suitable option for all users. It should not be regarded
as a one-size-fits-all solution or as a sufficient substitute for an easy online cancellation process.

Our results shows that more European news sites provide information relating to auto-renewal and cancelling than the US sites. On the other hand, the US news sites display more exit surveys and special offers to stay (Figure~\ref{fig:EuropeVusa}). Exit surveys can provide meaningful input to businesses on what they should improve on. However this should be optional so the users do not feel forced to justify their decisions. In addition, forcing users may lead to less truthful answers. 
European news sites require fewer clicks to cancel in comparison to their American counterparts (Figure~\ref{fig:useuropeclick}). Differences were noted between some countries. For example, the Dutch subscriptions were all paid via direct debit and displayed their special offers on their subscription management page as opposed to subsequent pages after pressing cancel. Certain German news websites used cookie paywalls, which was not encountered in any of the other countries' news sites.
While our results show differences across jurisdictions, we cannot precisely attribute these changes to the differences in regulations. Other factors, including market dynamics, may have played a role.

To conclude this section, we would like to discuss possible paths to improve the user experience related to subscriptions and cancellations. 
From a regulatory perspective, we echo the three-fold route that Busch suggests \cite{busch2022updating}: (1) clear choices at the beginning of the subscription process, (2) reminders about auto-renewals and prolonged inactive subscriptions, and (3) a simple and unambiguous cancellation protocol.
The newly introduced FCCA regulation in Germany
prescribes the look and placement of the two-step termination button. These precise requirements may make it harder to circumvent regulations, but in the long term they may potentially suffer from being too \textit{technology-specific}, and may need to be updated.
Given the variety and complexity of cancellation flows observed in our study, a combination of high-level principles and specific regulations may offer the best tradeoff.

\subsection{Limitations}
\label{subsec:limitations}
The news sites that we investigated in this study are popular, and therefore may be subject to higher scrutiny and may be less likely to use manipulative design. While a random and varied list of newspapers could uncover other practices, studying popular websites has the advantage of studying the experience faced by millions of users.
Our study is based on a snapshot of the subscription and cancellation processes at the start of 2023. These practices may change in time due to upcoming regulations, user feedback or for other reasons\footnote{Indeed, when attempting to cancel a New York Times subscription in December 2023, one of the authors discovered that online cancellation was no longer an option, despite it being available during our data collection in January/February 2023.}. Moreover, businesses that use A/B testing or personalization might offer alternative cancellation options to other users, which may not be captured in our recordings.
Due to time limitations, we only signed up for weekly or monthly subscriptions, leaving out yearly subscriptions. 

We opted to focus on desktop subscriptions---as opposed to mobile---because our preliminary analysis of the Android and iOS subscription/cancellation flows resulted in highly uniform experiences (see Figure~\ref{fig:Android-Cancel} in Appendix~\ref{app:B}). As a result we decided to focus on desktop subscriptions to observe differences across websites and locations.

A more thorough and balanced study would have seen our American personas subscribing to all European sites, but this would have almost doubled the subscriptions made in the study.


Screen captures were recorded and coded by the lead (junior) researcher, whose main language was English, in order to identify dark patterns. To prevent potential biases and mistakes, the results were reviewed and discussed in several meetings with three senior researchers in the fields of law, computer science and HCI.
For non-English websites, we used Google Chrome's internal translator, and Gmail's translation option for emails. When these translators failed, we manually copied and pasted text to Google Translate, or used a smartphone app to translate non-English text.
While the accuracy of these translations is generally high, there may be minor mistranslations.
In a few cases where we doubted the translation, we confirmed the translation with a native speaker.

The parameters that we use to characterize user experiences, such as the number of clicks, do not fully capture all difficulties posed by the cancellation interfaces. For example, in certain cases, locating a cancellation link hidden within a maze of menus may be a larger challenge than clicking multiple buttons.
For instance, several times the cancellation link was located at the bottom of the page, which could have been missed if the user did not scroll down.
Studying user experience metrics that consider the mental effort caused by these designs could help us understand the challenges of subscription and cancellation interfaces better.
While we carefully inspected the websites for cancellation links, some websites may have more than one way to cancel subscriptions. In such websites we might not have taken the shortest path to cancellation.

\subsection{Future Work}
We limited ourselves to five geographic locations and desktop-only subscriptions, to stay within the study budget. Future studies may use our methods to study news websites from different countries, including the global south.
Another direction for further research would be to study different types of subscriptions, such as streaming services, online games, and physical deliveries (e.g., cosmetic products, cleaning supplies, groceries, clothes).
Future work may also focus on different device form factors such as mobile.
These studies may reveal novel forms of manipulative designs, or best practices that are not employed by the news websites that we studied.

A longitudinal study, considering changes over time, might also shed light on the effects of regulatory changes or industry trends. Our findings and dataset may serve as a snapshot for such studies.
Buying multiple subscriptions for the same site could identify A/B testing and help investigate the consistency of the subscription and cancellation process over multiple attempts.
A user study could investigate what users find particularly challenging in subscription and cancellation flows, and what dark patterns are particularly effective in preventing cancellations against users' will. Such studies may make recommendations for principles and building blocks for optimal subscription and cancellation flow designs.
Additionally, the emails sent by the subscription services offer an opportunity for further research, exposing the tactics employed to win back unsubscribed users.

In line with prior work, our work is mostly based on analyzing design patterns in isolation---except a few cases including Bait and Switch (Figure~\ref{fig:RP+ColourSwitch}),
where two dialogues in a given cancellation flow were compared.
Recent proposals to \textit{temporally} analyze dark patterns may provide more insights into the compounding effects of dark patterns within cancellation flows~\cite{gray2023temporal}.
Finally, while we used the dark pattern taxonomy of Mathur et al.~\cite{mathur2019dark}, future research may consult the comprehensive ontology presented in Gray et al.'s concurrent work~\cite{gray2023ontology} for a more nuanced, layered and standardized categorization.

\textbf{Ethics}
During our study we took appropriate measures to avoid interfering with the regular operation of the businesses to which we subscribed.
By using simulated personas with fictitious names and addresses, we aimed to protect the identity and security of our research team.
Only in a few cases, were we required to provide the real name of the lead author, because the name on the credit card was needed.
Since we subscribed to popular news sites with thousands of subscribers, we believe any negative effect due to our use of the simulated personas will be minimal.
Moreover, using simulated personas~\cite{roesner2012detecting,barford2014adscape} and mystery shopping~\cite{turner2012mystery} are established methods for evaluating business practices.
We only engaged with real customer representatives when we were required to call to cancel (four times in total), during which we focused solely on cancelling subscriptions quickly. These conversations were brief and resembled typical interactions with potential customers. Given our research's focus on business practices, we did not disclose that these calls were part of a research study. We did not record our calls with the representatives.

%% file: sections/conclusion.tex
\section{Conclusion}
\label{sec:con}

Subscriptions provide businesses a steady stream of revenue, but they might unfairly challenge users with insufficient information during subscription and obstacles during cancellation.
To the best of our knowledge, our study is the first attempt to characterize the difficulties encountered by users when trying to cancel online subscriptions across a range of news websites, from multiple vantage points.
Our results show that the majority (63/67) of the subscriptions could be cancelled online, with a few exceptions where we were required to make a phone call. Overall, the low number of websites requiring users to cancel via phone is a positive finding and may be due to recent regulations and enforcement actions, or raised user awareness around dark patterns. On the other hand, we could cancel all subscriptions online only from two of the five locations we studied: Germany and California. Notably these two locations have relatively stricter regulations on subscriptions compared to other studied locations within Europe and the US, respectively.
Our findings indicate that online cancellation processes are not without their share of barriers and complexities.
We documented cases of dark patterns in cancellation flows including \emph{Misdirection}, \emph{Obstruction} and \emph{Forced Action}.
Specifically, we uncovered long and arduous cancellation flows, misleading button placements, compulsory exit surveys, and an interface that forced users to enter a certain phrase.
Studying sign-up flow of the subscriptions, we found several cases where the users are not fully informed about the terms of the subscriptions such as recurring payments.
Our analysis showed that compared to their European counterparts, American news sites require more clicks to cancel, they request more information when subscribing, and they are more likely to present mandatory exit surveys.
While we believe that recent and upcoming regulatory updates rightfully focus on preventing problematic practices related to subscriptions, their success depends on deterrence and effective enforcement. In the meantime, we believe businesses
should focus on gaining and retaining users' trust instead of short-term gains.

%% file: sections/appendices.tex
\appendix

\section{Subscription Study Protocol}
\label{app:A}
To make sure that the subscription process ran smoothly and without mistakes, we compiled a standard protocol. This prevented any mistakes in the process, for example forgetting to change VPN location to the persona's location, and also save time on entering the details during account sign-ups.

\begin{enumerate}
    \item Before subscribing to a website, have your persona's details ready for copying and pasting. The card details should also be on same page, along with passwords needed for accounts.
    \item Another spreadsheet should be created  containing all media links and details should also be open.
    \item For each individual persona, set up a Chrome account. 
    \item When subscribing from each individual persona's country, turn on VPN for their location and open their Chrome account.
    \item Switch to the persona's Chrome account, click on the media link in the spreadsheet for each newspaper and start screen recording.
    \item Scroll down through the website to document the parameters on all pages you click through.
    \item Record in original language.
    \item Go through the process of subscribing to the account using Chrome's autofill to save time and stop recording when done.
    \item When presented with different subscription terms such as weekly, monthly and yearly, choose the shortest term.
    \item When presented with different subscription or contribution prices for the same term, choose the lowest non-zero amount.
    \item Note details under each column in the spreadsheet.
    \item Repeat for cancelling subscriptions.
    \item When presented with an exit survey, try submitting it without choosing an option to determine whether it is mandatory or not.
    \item When phoning to cancel, have subscription details ready for verification.
    \item Save recordings for subscribing and cancelling in separate folders.
\end{enumerate}

\section{Cancelling The New York Times Subscription Via Android}
\label{app:B}
\begin{figure*}
  \centering 
  \frame{\subfloat[]{\includegraphics[width=0.2\textwidth]{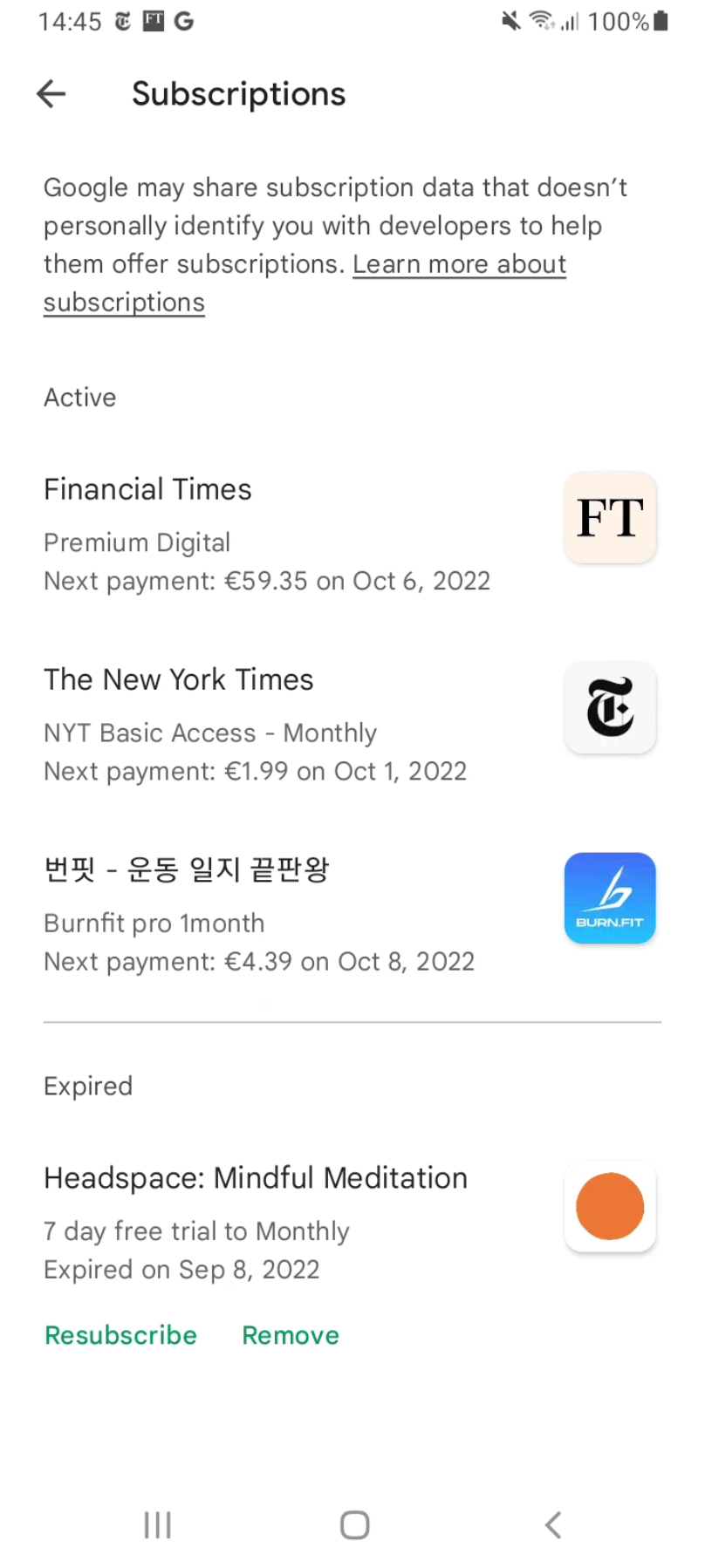}}} 
  \frame{\subfloat[]{\includegraphics[width=0.2\textwidth]{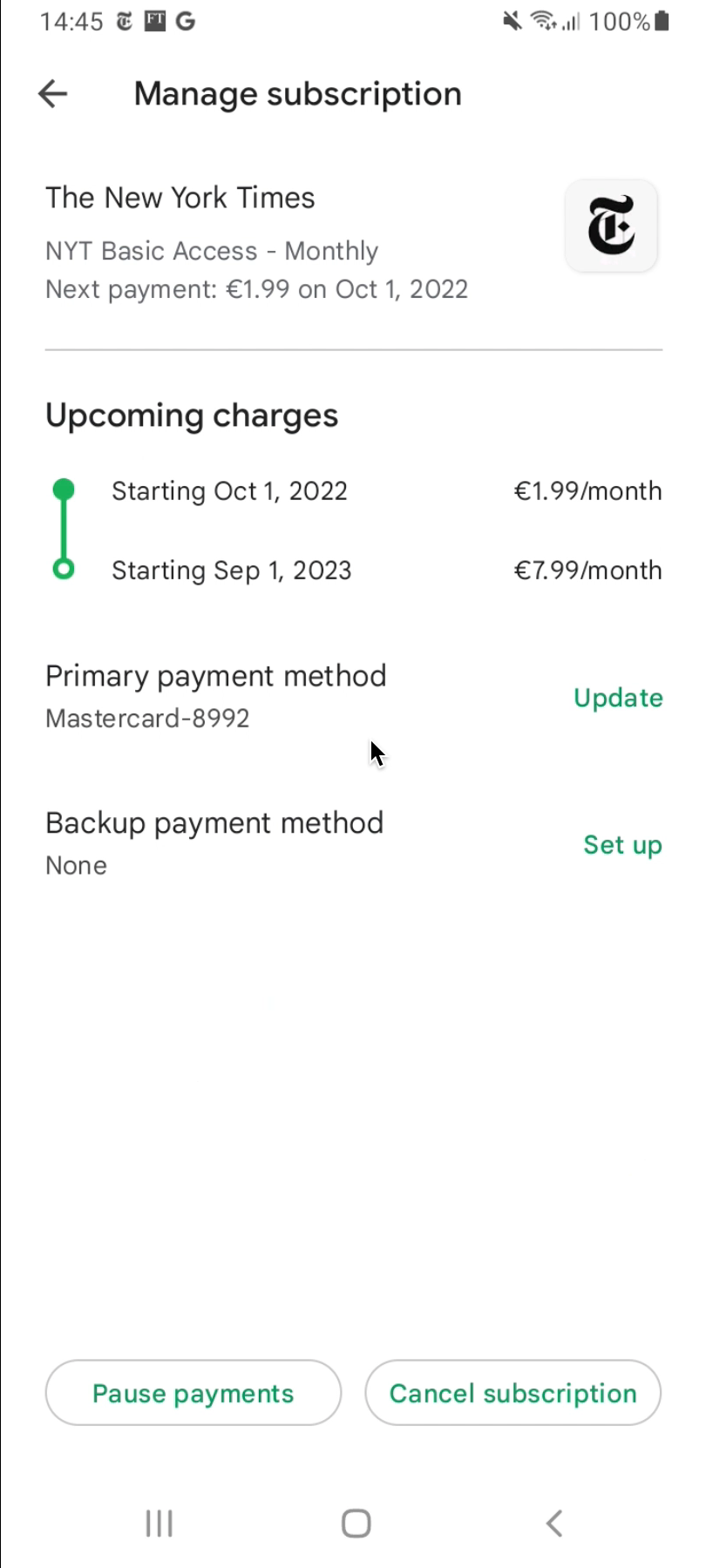}} }
  \frame{\subfloat[]{\includegraphics[width=0.2\textwidth]{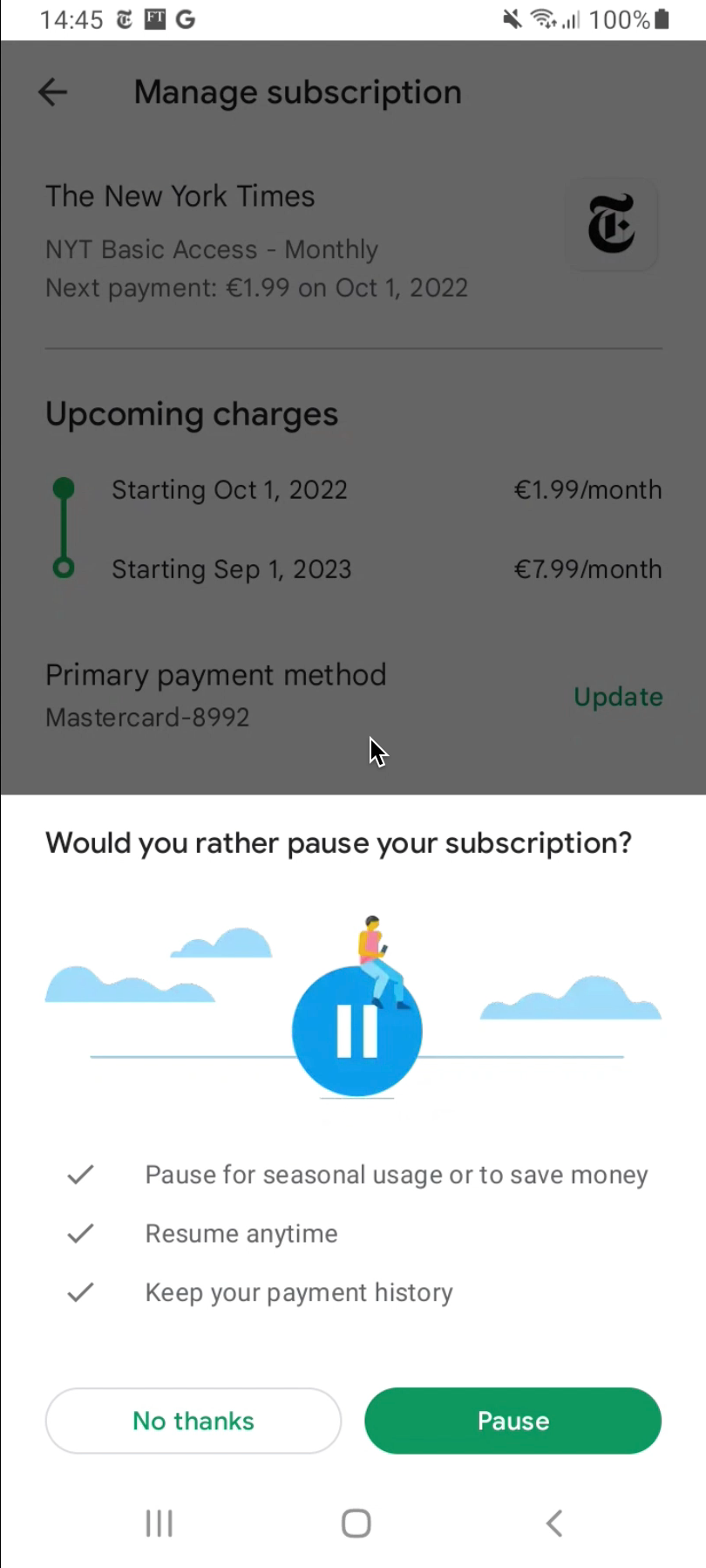}}}\\
\frame{\subfloat[]{\includegraphics[width=0.2\textwidth]{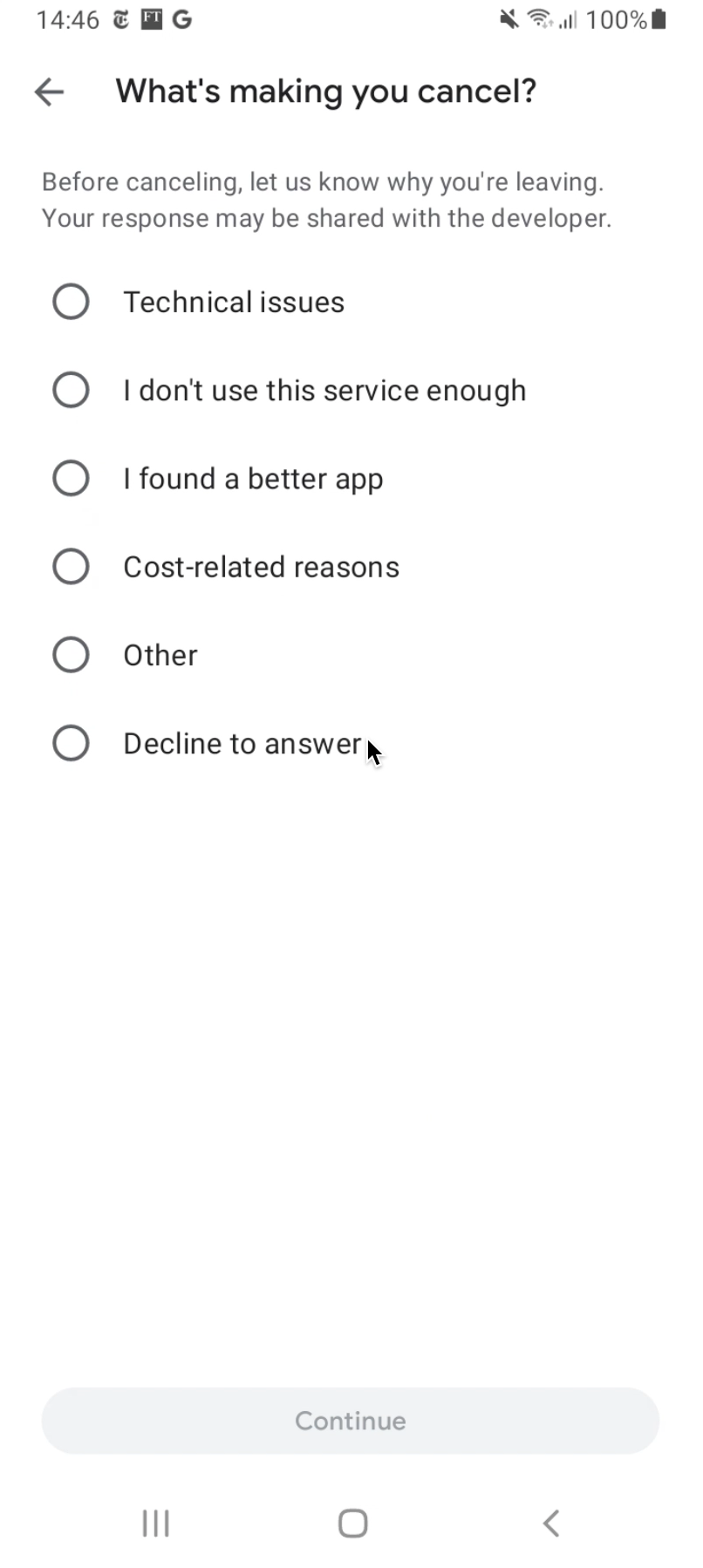}}} 
  \frame{\subfloat[]{\includegraphics[width=0.2\textwidth]{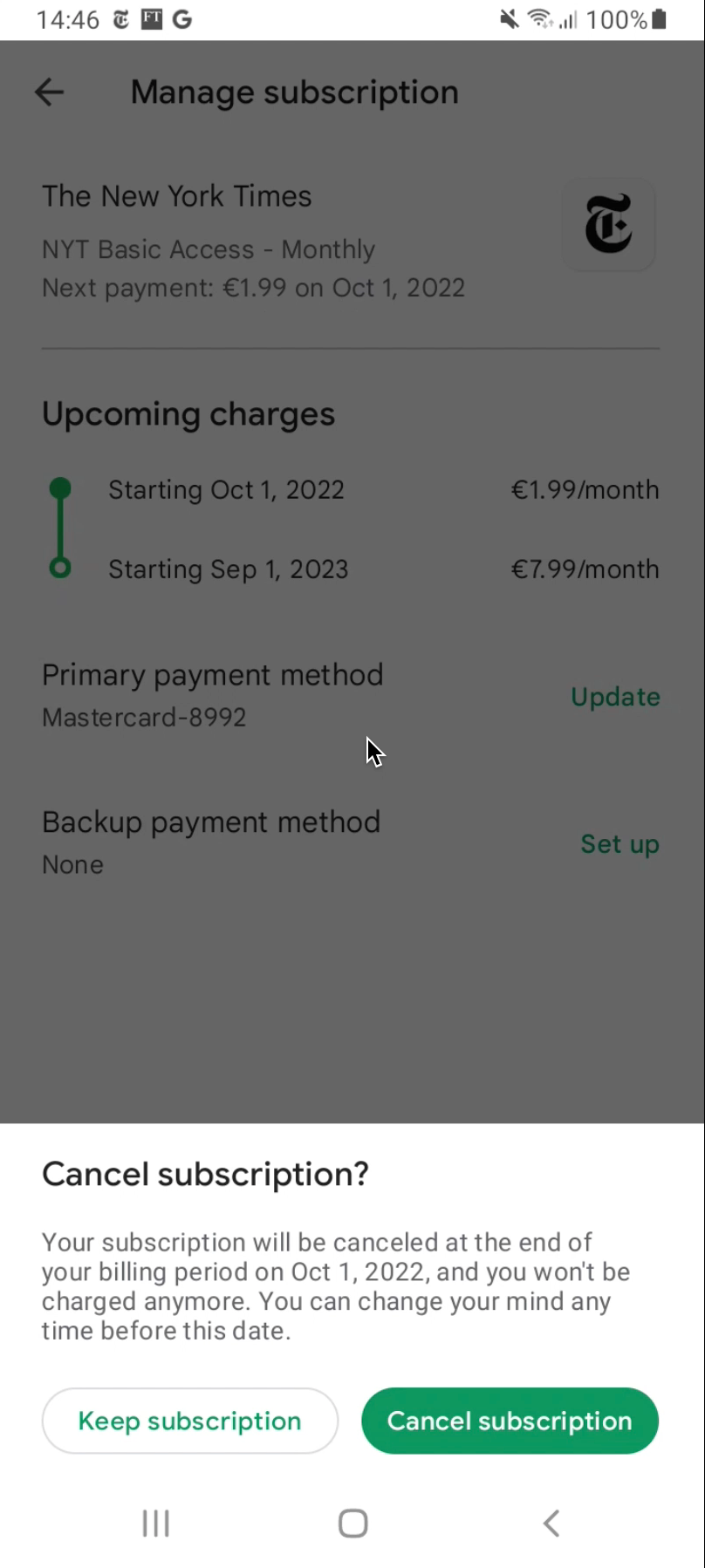}} }
  \frame{\subfloat[]{\includegraphics[width=0.2\textwidth]{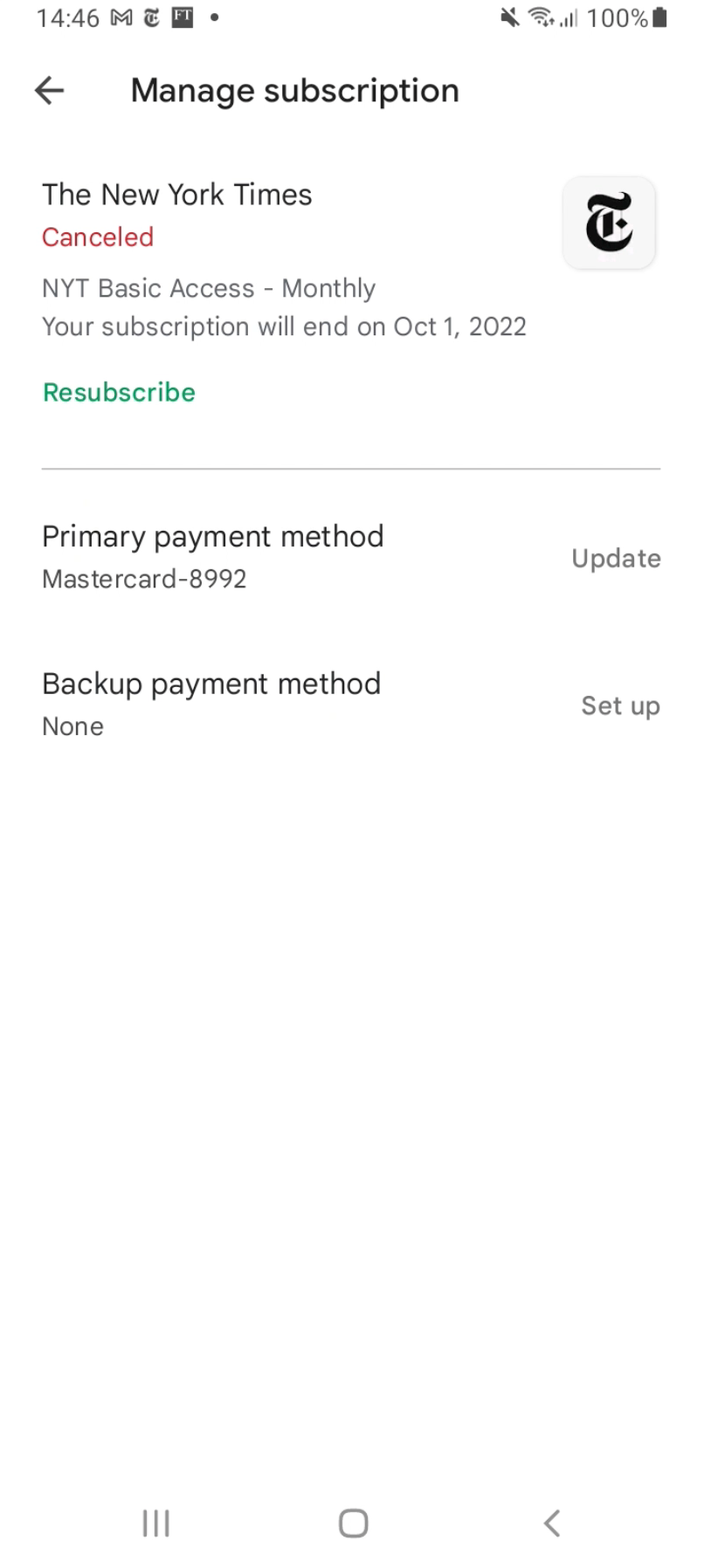}}}
  \caption{Android cancellation process for The New York Times app. The process takes five clicks and includes a mandatory survey. }
  \label{fig:Android-Cancel}
\end{figure*}

As mentioned in Section~\ref{sec:intro}, we performed a preliminary study where we looked at different subscription types and platforms before deciding to focus solely on news sites. In our investigations into mobile app subscriptions, we used an Android phone and found that because apps were downloaded via the Google store, the subscription cancellation process was uniform. Figure~\ref{fig:Android-Cancel} shows the standard Google cancellation process, which includes a mandatory survey and a notice that asks if you would rather pause your subscription instead. Managing subscriptions through one location, such as Google Play or Apple's App Store, does make the cancellation process easier.

\onecolumn
\section{Additional Results for Subscriptions to US News Sites}
\label{app:C}

\begin{table*}[h]

\resizebox{\textwidth}{!}{%
\begin{tabular}{@{}lrrrrrrrrrrrrrrr@{}}

\toprule

\textbf{US Media} & \multicolumn{3}{c}{\textbf{Texas USA}}  & \multicolumn{3}{c}{\textbf{California USA}} & \multicolumn{3}{c}{\textbf{Germany}} & \multicolumn{3}{c}{\textbf{Netherlands}}  & \multicolumn{3}{c}{\textbf{United Kingdom}} \\
\cmidrule(rl){2-4} \cmidrule(rl){5-7}\cmidrule(rl){8-10} \cmidrule(rl){11-13}\cmidrule(rl){14-16} 
& \multicolumn{1}{c}{\begin{tabular}[c]{@{}c@{}}Clicks to  \\
 Subscribe\end{tabular}} & \multicolumn{1}{c}{\begin{tabular}[c]{@{}c@{}}Clicks to \\ Cancel\end{tabular}} & \multicolumn{1}{c}{\begin{tabular}[c]{@{}c@{}} Extra \\ Clicks\end{tabular}} & \multicolumn{1}{c}{\begin{tabular}[c]{@{}c@{}}Clicks to\\  Subscribe\end{tabular}} & \multicolumn{1}{c}{\begin{tabular}[c]{@{}c@{}}Clicks to\\  Cancel\end{tabular}} & \multicolumn{1}{c}{\begin{tabular}[c]{@{}c@{}} Extra \\ Clicks\end{tabular}} & \multicolumn{1}{c}{\begin{tabular}[c]{@{}c@{}}Clicks to  \\  Subscribe\end{tabular}} & \multicolumn{1}{c}{\begin{tabular}[c]{@{}c@{}}Clicks to\\  Cancel\end{tabular}} & \multicolumn{1}{c}{\begin{tabular}[c]{@{}c@{}} Extra\\  Clicks\end{tabular}} & \multicolumn{1}{c}{\begin{tabular}[c]{@{}c@{}}Clicks to  \\  Subscribe\end{tabular}} & \multicolumn{1}{c}{\begin{tabular}[c]{@{}c@{}}Clicks to \\ Cancel\end{tabular}} & \multicolumn{1}{c}{\begin{tabular}[c]{@{}c@{}}Extra\\  Clicks\end{tabular}} & \multicolumn{1}{c}{\begin{tabular}[c]{@{}c@{}}Clicks to  \\  Subscribe\end{tabular}} & \multicolumn{1}{c}{\begin{tabular}[c]{@{}c@{}}Clicks to \\ Cancel\end{tabular}} & \multicolumn{1}{c}{\begin{tabular}[c]{@{}c@{}}Extra \\ Clicks\end{tabular}} \\

\cmidrule(rl){2-4} \cmidrule(rl){5-7}\cmidrule(rl){8-10} \cmidrule(rl){11-13}\cmidrule(rl){14-16} 
The New York Times  & 4    & 7   & 3   & 4  & 7    & 3   & 5   & 7  & 2  & 4  & 7  & 3 &5  & 7    & 2 \\
Wall Street Journal & 5  & *  & *  & 5  & 5  & 0  & 4  & 5 & 1   & 4  & 5  & 1 & 4 & * & * \\
Washington Post  & 4  & 6  & 2  & 4  & 6 & 2  & 3    & 6   & 3  & 3  & 6  & 3  & 3   & 6 & 3  \\
The Athletic & 2  & 6  & 4 & 2 & 6  & 4  & 2   & 6   & 4  & 2  & 6 & 4  & 2  & 6  & 4   \\
Substack  & 3  & 5   & 2   & 3   & 3   & 0   & 3   & 6  & 3  & 3  & 6  & 3  & 3  & 6 & 3\\
Medium  & 3  & 5   & 2   & 3  & 5   & 2  & 3  & 5  & 2  & 3  & 5 & 2 & 3 & 5 & 2 \\
The Daily Wire & 3  & 4  & 1  & 3   & 4  & 1  & 3  & 4  & 1  & 3 & 4 & 1  & 3  & 4 & 1  \\
Barrons   & 5   & 5  & 0  & 5   & 5   & 0  & *  & *  & *  & *  & *  & *  & * & *  & *   \\
Bloomberg Media & 3   & 10   & 7 & 3  & 10   & 7  & 3  & 10  & 7 & 3  & 10 & 7  & 3  & 10  & 7  \\
& \multicolumn{1}{l}{}  & \multicolumn{1}{l}{}  & \multicolumn{1}{l}{}  & \multicolumn{1}{l}{}  & \multicolumn{1}{l}{}  & \multicolumn{1}{l}{}  & \multicolumn{1}{l}{}  &\multicolumn{1}{l}{} & \multicolumn{1}{l}{} & \multicolumn{1}{l}{} & \multicolumn{1}{l}{}   & \multicolumn{1}{l}{}  & \multicolumn{1}{l}{}  & \multicolumn{1}{l}{}  & \multicolumn{1}{l}{} \\ \midrule
\textbf{Average}  & 3.5  & 6 & 2.6   & 3.5   & 5.6  & 2.1  & 3.3  & 6.1  & 2.9  & 3.1  & 6.1 & 3  & 3.3  & 6.3  & 3.1        \\ \bottomrule
\end{tabular}%
}
\caption{Number of clicks required to subscribe to \& cancel from American news sites by all five personas.}
\label{tab:clicks}
\end{table*}

\section{Additional Study Parameters}
\label{app:D}
\begin{table}[H]
\resizebox{0.8\columnwidth}{!}{%
\begin{tabular}{@{}lcl@{}}
\toprule
\textbf{Parameter}        & \textbf{Table} & \textbf{Description}                                                                                        \\ \midrule
Cancellation Phone Call Duration & --
   &
  \begin{tabular}[c]{@{}l@{}}When calling to cancel, how long does the \\ whole process take? \end{tabular} \\ \midrule
Other Barriers/Details   &   --             & \begin{tabular}[c]{@{}l@{}}What other barriers were there \\ to cancelling, particularly noting dark patterns?\end{tabular} \\ \midrule
Subscription Price       & 1              & \begin{tabular}[c]{@{}l@{}}What was the price of the \\ subscription?\end{tabular}         \\ \midrule
Trials &
  1 &
  \begin{tabular}[c]{@{}l@{}}If there is a trial offered, is it free or how \\ much is it reduced, and what is the trial period?\end{tabular} \\ \midrule
Trials Post-cancellation & --
   &
  \begin{tabular}[c]{@{}l@{}}If a trial period is offered, does the \\ trial continue after cancelling?\end{tabular} \\ \midrule
Payment Method           &    --           & \begin{tabular}[c]{@{}l@{}}Do you have to pay via credit\\ card or direct debit?\end{tabular}               \\ \midrule
Cookie Paywall           &      --          & \begin{tabular}[c]{@{}l@{}}Is there a cookie paywall on the \\ interface when subscribing?\end{tabular}     \\ \midrule
Auto-renew Button        &       --         & \begin{tabular}[c]{@{}l@{}}Is there a button you can toggle \\ to turn off auto-renewal?\end{tabular}       \\ \midrule
Auto-renew Acknowledgement Tick Box &--
   &
  \begin{tabular}[c]{@{}l@{}}Is there a tick box to acknowledge your \\ understanding of auto-renewal information?\end{tabular} \\ \midrule
Duration of Subscription & 1              & \begin{tabular}[c]{@{}l@{}}Is the subscription offered weekly, \\ monthly, yearly or other?\end{tabular}    \\ \midrule
No. Subscribers          & 1              & How many subscribers according to [74]?                                                                     \\ \midrule
Tranco Rank              & 1              & What rank is news site according to Tranco?   \\ \bottomrule                                                             
\end{tabular}%
}
\caption{List of additional parameters recorded when subscribing and cancelling each account. The ``Table'' column shows
in which table the corresponding results can be found. 
}
\label{tab:additional-params}
\end{table}